\documentclass{sig-alternate-10}
\usepackage{balance}

\usepackage[dvipsnames,table]{xcolor}
\usepackage{xspace}

\usepackage{afterpage}

\usepackage{tikz}
\usetikzlibrary{calc}
\usetikzlibrary{shapes.misc}
\usepackage{pgfplots}
\usepackage{pgfplotstable}
\pgfplotsset{compat=1.8}
\usepgfplotslibrary{statistics}
\usepgfplotslibrary{colorbrewer}
\pgfplotsset{cycle list/Dark2-6}
\pgfplotsset{cycle list/Dark2-7}

\usepackage{booktabs}
\usepackage{url}

\newcommand{\cB}{\ensuremath{\mathcal{B}}}
\newcommand{\cE}{\ensuremath{\mathcal{E}}}
\newcommand{\cH}{\ensuremath{\mathcal{H}}}
\newcommand{\cI}{\ensuremath{\mathcal{I}}}
\newcommand{\cL}{\ensuremath{\mathcal{L}}}
\newcommand{\cO}{\ensuremath{\mathcal{O}}}
\newcommand{\cV}{\ensuremath{\mathcal{V}}}

\newcommand{\vars}{\text{vars}}

\newcommand{\ptime}{\textsc{Ptime}\xspace}
\newcommand{\np}{\textsc{NP}\xspace}
\newcommand{\sigmatwo}{\ensuremath{\Sigma_2^P}\xspace}
\newcommand{\pspace}{\textsc{Pspace}\xspace}

\newcommand{\dbpediaTwel}{\texttt{DBpedia9/12}\xspace}
\newcommand{\dbpediaThir}{\texttt{DBpedia13}\xspace}
\newcommand{\dbpediaFour}{\texttt{DBpedia14}\xspace}
\newcommand{\dbpediaFift}{\texttt{DBpedia15}\xspace}
\newcommand{\dbpediaSixt}{\texttt{DBpedia16}\xspace}
\newcommand{\lgdThir}{\texttt{LGD13}\xspace}
\newcommand{\lgdFour}{\texttt{LGD14}\xspace}
\newcommand{\biopThir}{\texttt{BioP13}\xspace}
\newcommand{\biopFour}{\texttt{BioP14}\xspace}
\newcommand{\swdfThir}{\texttt{SWDF13}\xspace}
\newcommand{\rkbeFour}{\texttt{BritM14}\xspace}

\newcommand{\biomedThir}{\texttt{BioMed13}\xspace}
\newcommand{\wikiSeven}{\texttt{WikiData17}\xspace}

\newcommand{\cq}{\text{CQ}\xspace}
\newcommand{\cqs}{\text{CQs}\xspace}
\newcommand{\cpf}{\text{CPF}\xspace}
\newcommand{\cqf}{\text{CQ}$_\text{F}$\xspace}
\newcommand{\cqfo}{\text{CQ}$_\text{OF}$\xspace}
\newcommand{\cqof}{\cqfo}
\newcommand{\aof}{\textsf{AOF}\xspace}

\newcommand{\sparql}{SPARQL\xspace}

\newcommand{\select}{\textsf{Select}\xspace}
\newcommand{\ask}{\textsf{Ask}\xspace}
\newcommand{\construct}{\textsf{Construct}\xspace}
\newcommand{\describe}{\textsf{Describe}\xspace}

\newcommand{\distinct}{\textsf{Distinct}\xspace}
\newcommand{\limit}{\textsf{Limit}\xspace}
\newcommand{\offset}{\textsf{Offset}\xspace}
\newcommand{\orderby}{\textsf{Order By}\xspace}

\newcommand{\filter}{\textsf{Filter}\xspace}
\renewcommand{\and}{\textsf{And}\xspace}
\newcommand{\union}{\textsf{Union}\xspace}
\newcommand{\opt}{\textsf{Opt}\xspace}
\newcommand{\graph}{\textsf{Graph}\xspace}
\renewcommand{\exists}{\textsf{Exists}\xspace}
\newcommand{\notexists}{\textsf{Not Exists}\xspace}
\newcommand{\minus}{\textsf{Minus}\xspace}

\newcommand{\bind}{\textsf{Bind}\xspace}
\newcommand{\countop}{\textsf{Count}\xspace}
\newcommand{\avgop}{\textsf{Avg}\xspace}
\newcommand{\minop}{\textsf{Min}\xspace}
\newcommand{\maxop}{\textsf{Max}\xspace}
\newcommand{\sumop}{\textsf{Sum}\xspace}
\newcommand{\groupby}{\textsf{Group By}\xspace}
\newcommand{\having}{\textsf{Having}\xspace}

\newcolumntype{R}{>{\kern\tabcolsep}r<{\kern\tabcolsep}@{}}

\newtheorem{theorem}{Theorem}[section]
\newtheorem{definition}[theorem]{Definition}
\newtheorem{example}[theorem]{Example}

\begin{document}

\title{An Analytical Study of Large SPARQL Query Logs}

\numberofauthors{3}

\author{
\alignauthor
Angela Bonifati\titlenote{Partially supported by CNRS Mastodons MedClean.}\\
       \affaddr{Lyon 1 University}\\
       \affaddr{Lyon, France}\\
\alignauthor
Wim Martens\\
       \affaddr{University of Bayreuth}\\
       \affaddr{Bayreuth, Germany}\\
\alignauthor
Thomas Timm\\
       \affaddr{University of Bayreuth}\\
       \affaddr{Bayreuth, Germany}\\
}

\maketitle

\begin{abstract}
  With the adoption of RDF as the data model for Linked Data and the
  Semantic Web, query specification from end-users has become more and
  more common in SPARQL endpoints. In this paper, we conduct an
  in-depth analytical study of the queries formulated by end-users and
  harvested from large and up-to-date query logs from a wide variety
  of RDF data sources.
  As opposed to previous studies, ours is the first assessment on a
  voluminous query corpus, spanning over several years and covering
  many representative SPARQL endpoints. Apart from the syntactical
  structure of the queries, that exhibits already interesting results
  on this generalized corpus, we drill deeper in the structural
  characteristics related to the graph- and hypergraph representation
  of queries. We outline the most common shapes of queries when
  visually displayed as pseudographs, and characterize their
  (hyper-)tree width. Moreover, we analyze the evolution of queries
  over time, by introducing the novel concept of a streak, i.e., a
  sequence of queries that appear as subsequent modifications of a
  seed query.  Our study offers several fresh insights on the already
  rich query features of real SPARQL queries formulated by real users,
  and brings us to draw a number of conclusions and pinpoint future
  directions for SPARQL query evaluation, query optimization, tuning,
  and benchmarking.
\end{abstract}

\section{Introduction}

As more and more data is exposed in RDF format, we are witnessing a compelling need from end-users to formulate more or less sophisticated queries on top of this data. SPARQL endpoints
are increasingly used to harvest query results from available RDF data repositories.  But how do these end-user queries look like? As opposed to RDF data, which can be easily obtained under the form of dumps (DB\-pedia and Wikidata dumps~\cite{wikidata, dbpediaDS, VrandecicK14}), query logs are often inaccessible, yet hidden treasures to understand the actual usage of these data. In this paper, we investigate a large corpus of query logs from different SPARQL endpoints,
which spans over several years (2009--2017). In comparison to previous studies on real SPARQL queries~\cite{Moller-wsc2010, DBLP:journals/corr/abs-1103-5043, PicalausaV-swim11, lsq-pd-2015, HanFZWRJ-webdb16}, which typically\footnote{The exception is \cite{HanFZWRJ-webdb16}, where logs from the Linked SPARQL Queries Dataset (LSQ) were studied, combining data from four sources (from 2010 and 2014) that we also consider.} investigated query logs of a single source,
we consider a multi-source query corpus that is two orders of magnitude larger. Furthermore, our analysis goes significantly deeper. In particular, we are the first to do a large-scale analysis on the topology of queries, which has seen significant theoretical interest in the last decades (e.g., \cite{ChekuriR-icdt97,GottlobGLS-pods16,GottlobLS-jcss02}) and is now being used for state-of-the-art structural decomposition methods for query optimization \cite{AbergerTOR-sigmod16,AbergerTOR-icdew16,KalinskyEK-edbt17}.
As a consequence, ours is the first analytical study on real (and most recent) SPARQL queries from a variety of domains reflecting the recent advances in theoretical and system-oriented studies of query evaluation.

Our paper makes the following contributions.  Apart from classical measures of syntactic properties of the investigated queries, such as their keywords, their number of triples and operator distributions, which we apply to our new corpus, we also mine the usage of projection in queries and subqueries in the various datasets. Projection indeed is the cause of increased complexity (from \ptime to \np-Complete) of the following central decision problem in query evaluation~\cite{ChandraM-stoc77, BarceloPS-pods15, LeteliePPS-tods13}: Given a conjunctive query $Q$, a database $D$, and a candidate answer $a$, is $a$ an answer of $Q$ on $D$?

We then proceed by considering queries
under their graph- and hypergraph structures. Such structural aspects of queries have been investigated in theory for over two decades \cite{GottlobGLS-pods16} since they can indicate when queries can be evaluated efficiently. Recently, several studies on new join algorithms leverage the hypergraph structure of queries in the contexts of relational- and RDF query processing \cite{AbergerTOR-sigmod16,KalinskyEK-edbt17}. Theoretical research in this area traditionally focused on \emph{conjunctive queries (CQs)}. For CQs, we know that tree-likeness of their structure leads to polynomial-time query evaluation \cite{GottlobGLS-pods16}. For larger classes of queries, the topology of the graph of a query is much less informative. For instance, if we additionally allow SPARQL's \opt operator, evaluation can be \np-complete even if the structure is a tree \cite{BarceloPS-pods15}.  For this reason, we focus our structural study on CQ-like queries.\footnote{We do consider extensions with \filter and \opt, but only those for which we know that tree-likeness of their graph ensures the existence of efficient evaluation algorithms.}
We develop a shape classifier for such queries and identify their most occurring shapes. Interestingly enough, these queries have quite regular shapes. The overwhelming majority of the queries is acyclic (i.e., tree- or forest-shaped). We discovered that the cyclic queries mostly consist of a central node with simple, small attachments (which we call \emph{flower}). In terms of tree- and hypertreewidth, we discovered that the cyclic queries have width two, up to a few exceptions with width three.

At this point we should make a note about interpretation of our results. Even though almost all CQ-like queries have (hyper-)treewidth one, we do not want to claim that queries of larger treewidth are not important in practice.  The overwhelming majority of the queries we see in the logs are small and simple and we believe this to be typical for SPARQL endpoint logs. For instance, the majority ($>$55\%) of the queries in our logs only use one triple. One of our data sets, \wikiSeven is not a SPARQL endpoint log and we see throughout the paper that it has completely different characteristics.

In order to gauge the performances of cyclic and acyclic queries from a
practical viewpoint, we have run a comparative analysis of chain and cycle
queries synthetically generated with an available graph and query workload generator \cite{BaganBCFLA17}. This experiment showed different behaviors of
SPARQL query engines, such as Blazegraph and PostgreSQL with query workloads of CQs of
increasing sizes (intended as number of conjuncts). It also lets us grasp
a tangible difference between chain and cycle queries in either query
engine, this difference being more pronounced for PostgreSQL.
We may interpret this result as a lack of maturity of practical query
engines for
cyclic queries, thus motivating the need of specific query optimization techniques for such queries as in \cite{AbergerTOR-sigmod16,KalinskyEK-edbt17}.

Finally, we deal with the problem of identifying sequences of similar queries in the query logs. These queries are then classified as gradual modifications of a seed query, possibly by the same user. We measure the length of such streaks in three log files from DBpedia.  We conclude our study with insights on the impact of our analytical study of large SPARQL query logs on query evaluation, query optimization, tuning, and benchmarking.

\noindent {\bf Related Work.} Whereas several previous studies have focused on the analysis of real SPARQL queries, they have mainly looked at statistical features of the queries, such as occurrences of triple patterns, types of queries, query fragments and well-designed patterns~\cite{Moller-wsc2010, DBLP:journals/corr/abs-1103-5043, lsq-pd-2015, HanFZWRJ-webdb16}. The only early study that investigated the relationship between structural features of practical queries and query evaluation complexity has been presented in~\cite{PicalausaV-swim11}. However, they focus on a limited corpus (3M queries from DBpedia 2010) and in that sense their findings cannot be generalized. Our work moves onward by precisely characterizing the occurrences of conjunctive and non-conjunctive patterns under the latest complexity results, by performing an accurate shape analysis of the queries under their (hyper-)graph representation and introducing the evolution of queries over time.
USEWOD and DBpedia datasets have also been considered in
\cite{AriasFMF-corr11}.
It takes into account the log files from DBpedia and
SWDF reaching a total size of 3M. They mainly investigate the number of
triples and joins in the queries. Based on the observation of
\cite{NeumannW-vldbj10} that typically SPARQL graph patterns are typically
chains or star-shaped, they also look at their occurrences. They found very
scarce chains and high coverage of almost-star-shaped graph patterns, but
they do not characterize the latter. To the best of our knowledge, we are
the first to carry out a comprehensive shape analysis on such a large and diverse corpus of
SPARQL queries.

 \section{Data Sets}

Our data set has a total of $180,653,910$ queries, which were obtained
as follows. We obtained the 2013--2016 USEWOD query logs, DBPedia
query logs for 2013, 2014, 2015 and 2016 directly from
Openlink\footnote{\url{http://www.openlinksw.com}}, the 2014 British
Museum query logs from LSQ\footnote{\url{http://aksw.github.io/LSQ/}},
and we crawled the user-submitted example queries from
WikiData\footnote{\url{https://www.wikidata.org/wiki/Wikidata:SPARQL_query_service/queries/examples}}
in February 2017. These log files are associated with 7 different data
sources from various domains: DBpedia, Semantic Web Dog Food (SWDF),
LinkedGeoData (LGD), BioPortal (BioP), OpenBioMed (Bio\-Med), British
Museum (BritM), and WikiData.

\begin{table}[tb]
  \begin{tabular}{@{}lrrr@{}}
    \toprule
    \emph{Source}  &  \emph{Total \#Q} & \emph{Valid \#Q} & \emph{Unique
\#Q}\\
    \midrule
    \dbpediaTwel & 28,534,301 & 27,097,467 & 13,437,966\\
    \dbpediaThir &  5,243,853 &  4,819,837 &  2,628,005\\
    \dbpediaFour & 37,219,788 & 33,996,480 & 17,217,448\\
    \dbpediaFift & 43,478,986 & 42,709,778 & 13,253,845\\
    \dbpediaSixt & 15,098,176 & 14,687,869 &  4,369,781\\
    \midrule
    \lgdThir     &  1,841,880 &  1,513,868 &    357,842\\
    \lgdFour     &  1,999,961 &  1,929,130 &    628,640\\
    \midrule
    \biopThir    &  4,627,271 &  4,624,430 &    687,773\\
    \biopFour    & 26,438,933 & 26,404,710 &  2,191,152\\
    \midrule
    \biomedThir  &    883,374 &    882,809 &     27,030\\
    \midrule
    \swdfThir    & 13,762,797 & 13,618,017 &  1,229,759\\
    \midrule
    \rkbeFour    &  1,523,827 &  1,513,534 &    135,112\\
    \midrule
    \wikiSeven   &        309 &        308 &       308 \\
    \midrule[1.5pt]
    Total        &180,653,910 & 173,798,237 & 56,164,661\\
 \end{tabular}
\caption{Sizes of query logs in our corpus.\label{tab:datasets}}
\end{table}

Table~\ref{tab:datasets} gives an overview of the analyzed query logs,
along with their main characteristics. Since we obtained logs for
DBpedia from different sources, we proceeded as follows. \dbpediaTwel
contains the DBpedia logs from USEWOD'13, which are query logs from
2009--2012. All other \texttt{DBpedia'X} sets contain the query logs
from the year \texttt{'X}, be it from USEWOD or from
Openlink.\footnote{We discovered that we received three log files
  from USEWOD as well as from Openlink, in the sense that only the hash
  values used for anonymisation were different. These duplicate log
  files were deleted prior to all analysis and are not taken into account in
  Table~\ref{tab:datasets}.}
We first cleaned the logs, since some contained entries that were
not queries (e.g., http
requests). In the following we only report on the actual \sparql
queries in the logs. For each of the logs, the table summarizes the total number of queries
(\emph{Total}) and the number of queries that we could parse using
Apache Jena 3.0.1 (\emph{Valid}). From the latter set, we removed
duplicate queries, resulting in the unique queries that we could parse
(\emph{Unique}) and on which we focus in the remainder of the
paper~\footnote{We report in a related appendix~\cite{ourAppendix} the results for the
\emph{Valid} corpus, containing duplicates.}.
In summary, our corpus of query logs contains the latest blend of USEWOD
and Openlink DBPedia query logs (the latter providing 51M more
queries in the period 2013-2016 than the USEWOD corpus), plus BritM and
Wikidata queries. We are not aware of other existing
studies on such a large and up-to-date corpus.
Finally, although the online WikiData example queries (Feb 13th, 2017) are a manually curated set,
there was one query that we could not parse.\footnote{The query was
  called ``Public Art in Paris'' and was malformed (closing braces
  were missing and it had a bad aggregate). It was still malformed on
  June 29th, 2017.}
In the total unique data set, 2,496,806 queries (4.47\%) do not have a
body. All these queries are \describe queries and almost exclusively
occur in \dbpediaFour -- \dbpediaSixt.

\section{Preliminaries}

We recall some basic definitions on RDF and SPARQL
\cite{PerezAG-tods09,PicalausaV-swim11}. We closely follow the
exposition of \cite{PicalausaV-swim11}.

\paragraph*{RDF}
RDF data consists of a set of triples $\langle s,p,o \rangle$ where we
refer to $s$ as \emph{subject}, $p$ as \emph{predicate}, and $o$ as
\emph{object}. According to the specification, $s$, $p$, and $o$ can
come from pairwise disjoint sets $\cI$ (\emph{IRIs}), $\cB$
\emph{blank nodes}, and $\cL$ \emph{literals} as follows: $s \in \cI
\cup \cB$, $p \in \cI$, and $o \in \cI \cup \cB \cup \cL$. For this
paper, the distinction between IRIs, blank nodes, and literals is not
important.

\paragraph*{\sparql}
For our purposes, a \emph{\sparql query $Q$} can be seen as a tuple of the form
\begin{center}
  (\emph{query-type}, \emph{pattern $P$}, \emph{solution-modifier}).
\end{center}
We now explain how such queries work conceptually.  The central
component is the \emph{Pattern $P$}, which contains patterns that are
matched onto the RDF data. The result of this part of the query is a
multiset of mappings that match the pattern to the data.

The \emph{solution-modifier} allows aggregation, grouping, sorting,
duplicate removal, and returning only a specific window (e.g., the
first ten) of the multiset of mappings returned by the pattern. The
result is a list $L$ of mappings.

The \emph{query-type} determines the output of the query. It is one of
four types: \select, \ask, \construct, and \describe. \select-queries
return projections of mappings from $L$. \ask-queries return a boolean
and answer true if the pattern $P$ could be matched.
\construct queries construct a new set of RDF triples based on the
mappings in $L$. Finally, \describe queries return a set of RDF triples that
describes the IRIs in $\cI$ and the blank nodes in $L$. The exact output of
\describe queries is implementation-dependent. Such queries are meant
to help users explore the data.
With respect to \cite{PicalausaV-swim11}, we allow more solution modifiers
and more complex patterns, as explained next.

\paragraph*{Patterns}

Let $\cV = \{?x, ?y, ?z, ?x_1, \ldots\}$ be an infinite set of
variables, disjoint from $\cI$, $\cB$, and $\cL$. As in SPARQL, we
always prefix variables by a question mark. A \emph{triple
  pattern} is an element of $(\cI \cup \cB \cup \cV) \times (\cI \cup \cV)
\times (\cI \cup \cB \cup \cL \cup \cV)$. A \emph{property path} is a regular
expression over the alphabet $\cI$. A \emph{property path pattern} is
an element of $(\cI \cup \cB \cup \cV) \times pp \times (\cI \cup \cB \cup \cL \cup
\cV)$, where $pp$ is a property path. A \emph{\sparql pattern} is an
expression generated from the following grammar:
\begin{center}
  \begin{tabular}{ll}
    $P ::=$ & $t \mid pp \mid Q \mid P_1$ \and $P_2 \mid P$ \filter $R$\\
    & $\mid P_1$ \union $P_2 \mid P_1$ \opt $P_2 \mid$ \graph $iv$ $P$
  \end{tabular}
\end{center}
Here, $t$ is a triple pattern, $pp$ is a property path pattern, $Q$ is
again a SPARQL query, $R$ is a so-called \emph{SPARQL filter
  constraint}, and $iv \in \cI \cup \cV$. We note that property paths
($pp$) and subqueries ($Q$) in the above grammar are new features
since SPARQL 1.1. SPARQL filter constraints $R$ are built-in
conditions which can have unary predicates, (in)equalities between
variables, and Boolean combinations thereof. We refer to the SPARQL
1.1 recommendation \cite{sparql11} and the literature
\cite{PerezAG-tods09}
for the precise syntax of
filter constraints and the semantics of SPARQL queries.  We write
$\vars(P)$ to denote the set of variables occurring in $P$.

We illustrate by example how our definition corresponds to real \sparql
queries. The following query comes from WikiData \cite{wikidata} (``Locations of archaeological sites'',
from \cite{wikidata}).
\begin{verbatim}
SELECT ?label ?coord ?subj
WHERE
{ ?subj wdt:P31/wdt:P279* wd:Q839954 .
  ?subj wdt:P625 ?coord .
  ?subj rdfs:label ?label filter(lang(?label)="en")
}
\end{verbatim}
The query uses the property path \texttt{wdt:P31/wdt:P279*}, literal
\texttt{wd:Q839954}, and triple pattern \texttt{?subj wdt:P625
  ?coord}. It also uses a filter constraint. In \sparql, the \and
operator is denoted by a dot (and is sometimes implicit in
alternative, even more succinct syntax).

Finally, we define conjunctive queries, which are a central class of
queries in database research and which we will build on in the
remainder of the paper. In the context of SPARQL, we define them as follows.
\begin{definition}\upshape
  A \emph{conjunctive query (\cq)} is a \sparql pattern that only uses
  the triple patterns and the  operator \and.
\end{definition}

\begin{table}[tb]
  \centering
\begin{tabular}{@{}r rr@{}}
  \toprule
  \emph{Element}  & \emph{Absolute} & \emph{Relative}\\
  \midrule
  \select   & 49,409,913 & 87.97\%\\
  \ask      &  2,789,420 &  4.97\%\\
  \describe &  2,578,311 &  4.49\%\\
  \construct&  1,386,908 &  2.47\%\\
  \midrule
  \distinct & 12,198,198 & 21.72\%\\
  \limit    &  9,545,249 & 17.00\%\\
  \offset   &  3,455,500 &  6.15\%\\
  \orderby  &  1,159,231 &  2.06\%\\
  \midrule
  \filter   & 22,547,561 & 40.15\%\\
  \and      & 15,863,942 & 28.25\%\\
  \union    & 10,465,706 & 18.63\%\\
  \opt      &  9,106,419 & 16.21\%\\
  \graph    &  1,519,899 &  2.71\%\\
  \notexists&    926,849 &  1.65\%\\
  \minus    &    766,380 &  1.36\%\\
  \exists   &      5,499 &  0.01\%\\
  \midrule
  \countop  &    320,035 &  0.57\%\\
  \maxop    &      3,660 &  0.01\%\\
  \minop    &      3,632 &  0.01\%\\
  \avgop    &        263 & $<$ 0.01\%\\
  \sumop    &         68 & $<$ 0.01\%\\
  \groupby  &    168,444 &  0.30\% \\
  \having   &     12,276 &  0.02\%\\
\end{tabular}
\caption{Keyword count in queries\label{tab:keywords}}
\vspace{-5mm}
\end{table}

\section{Shallow Analysis}

In this section we investigate simple syntactical properties of queries.

\subsection{Keywords}
A basic usage analysis of SPARQL features was done by counting the
keywords in queries. The results are in Table~\ref{tab:keywords}.\footnote{We also investigated the occurrence of other operators
(\textsf{Service}, \textsf{Bind}, \textsf{Assign}, \textsf{Data},
\textsf{Dataset}, \textsf{Values}, \textsf{Sample}, \textsf{Group
  Concat}), each of which appeared in less than 1\% of the queries. We
omit them from the table for succinctness.}

The first block in Table~\ref{tab:keywords} describes the type of queries.
In total, 87.97\% of the queries are \select-queries, 4.97\% are
\ask-queries, 4.59\% \describe queries, and 2.47\% \construct
queries. There are, however, tremendous differences between the data
sets. \biomedThir has less than 13\% \select-queries and almost 85\%
\describe-queries, whereas \lgdThir has 28\% \select-queries and 71\%
\construct-queries. Even within the same kind of data, we see
significant differences. \dbpediaSixt has 62\% \select-queries (and
34\% \describe-queries), whereas \dbpediaFift has 81.5\%
\select-queries and 11.5\% \ask-queries. The other DBpedia data sets
have over 87.5\% \select queries.

The second block in Table~\ref{tab:keywords} contains solution
modifiers, ordered by their popularity.\footnote{The remaining
  solution modifier, \textsf{Reduced}, was only found in 1.113
  queries.} Looking into the specific data sets, we see the following
things stand out. Almost all (97\%) of \rkbeFour queries use
\distinct. This is similar, but to a lesser extent in \biopThir (82\%)
and \biopFour (69\%). In DBPedia we again see significant
differences. From '12 to '16, we have 18\%, 8\%, 11\%, 38\%, and 8\%
of queries with \distinct respectively.

\limit is used most widely in \swdfThir (47\%) and \lgdFour
(41\%). The most prevalent data sets for queries with \offset are
\lgdFour (38\%), \lgdThir (13\%), and \dbpediaThir (12\%).

\orderby is used by far the most in WikiData (42\%), which may be due to the case
that their queries are intended to showcase the system and should
produce a nice output. Another reason may be that the other query logs
also contain the ``development process'' of queries: Users start by
asking a query and gradually refine it until they have the one they
want. (We come back to this in Section~\ref{sec:streaks}).

The third block has keywords associated to SPARQL algebra operators
that occur in the body. We see that \filter, \and, \union, and \opt
are quite common.\footnote{Conjunctions in SPARQL are actually denoted by
  ``.'' or ``;'' for brevity, but we group them under ``\and'' in this paper for
  readability.} The next commonly used operator is \graph but, looking
closer at our data, we see that 95\% of the queries using \graph
originate from \biopThir and \biopFour. In these logs, 80\% and 40\%
of the queries use \graph, respectively. The use of \filter ranges
from 61\% (\lgdFour) to 3\% or less (\biomedThir,
\biopThir).

The fourth block has aggregation operators. We were surprised that
these operators are used so sparsely, even though aggregates are only
supported since SPARQL 1.1 (March 2013) \cite{sparql11}. In all data sets, each of these
operators was used in 3\% or less of the queries, except for
\lgdFour (31\% with \countop) and \wikiSeven (30\% with \groupby). We see
a higher relative use of aggregation operators in \wikiSeven than in
the other sets, which we believe may be due to the fact that our
\wikiSeven set is not a query log. \wikiSeven is in fact a wiki page
that contains cherry-picked and user-submitted queries, some of which are meant to
highlight features of the Wikidata data set.

\begin{figure*}[t]
  \centering

\pgfplotstableread[col sep=comma]
{triples.csv}
{\datatableE}

\begin{tikzpicture}
\begin{axis}[
  ybar stacked,
    legend style={at={(1,0.5)}, anchor=west},
    reverse legend,
    ylabel near ticks,
    yticklabel=\pgfmathparse{100*\tick}\pgfmathprintnumber{\pgfmathresult}\,\%,
    bar width=0.7cm,
    x=1.0cm,
    xticklabels={
      \dbpediaTwel,
      \dbpediaThir,
      \dbpediaFour,
      \dbpediaFift,
      \dbpediaSixt,
      \lgdThir,
      \lgdFour,
      \biopThir,
      \biopFour,
      \biomedThir,
      \swdfThir,
      \rkbeFour,
      \wikiSeven
    },
    xtick={1, ..., 13},
    x tick label style={rotate=45,anchor=east},
    legend cell align=left
  ]
  \addplot+[NavyBlue!80!black,fill=NavyBlue,ybar] table[x=id, y=0] from {\datatableE};
  \addplot+[ProcessBlue!80!black,fill=ProcessBlue,ybar] table[x=id, y=1] from {\datatableE};
  \addplot+[SkyBlue!80!black,fill=SkyBlue,ybar] table[x=id, y=2] from {\datatableE};
  \addplot+[BlueGreen!80!black,fill=BlueGreen,ybar] table[x=id, y=3] from {\datatableE};
  \addplot+[LimeGreen!80!black,fill=LimeGreen,ybar] table[x=id, y=4] from {\datatableE};
  \addplot+[SpringGreen!80!black,fill=SpringGreen] table[x=id, y=5] from {\datatableE};
  \addplot+[Goldenrod!80!black,fill=Goldenrod] table[x=id, y=6] from {\datatableE};
  \addplot+[YellowOrange!80!black,fill=YellowOrange,ybar] table[x=id, y=7] from {\datatableE};
  \addplot+[Orange!80!black,fill=Orange,ybar] table[x=id, y=8] from {\datatableE};
  \addplot+[Red!50!black,fill=Red,ybar] table[x=id, y=9] from {\datatableE};
  \addplot+[BrickRed!50!black,fill=BrickRed,ybar] table[x=id, y=10] from {\datatableE};
  \addplot+[black,fill=Sepia,ybar] table[x=id, y=11] from {\datatableE};

  \legend{0, 1, 2, 3, 4, 5, 6, 7, 8, 9, 10, $11+$}
\end{axis}
\end{tikzpicture}

\vspace{-1mm}
\centering
\resizebox{\textwidth}{!}{
\begin{tabular}{@{}cccccccccccccc@{}}
\toprule
Datasets &  \dbpediaTwel & \dbpediaThir  &   \dbpediaFour &   \dbpediaFift &   \dbpediaSixt &  \lgdThir   &     \lgdFour  &     \biopThir &    \biopFour &     \biomedThir  &     \swdfThir   &     \rkbeFour   &    \wikiSeven  \\
\textit{S/A} & 99.15\% & 91.88\% & 95.38\% & 93.05\% & 63.99\% &
29.01\% & 97.47\% & 100\% & 99.69\% & 12.87\% & 96.14\% & 98.64\% & 99.68\%\\
\textit{Avg}$_{\#T}$ & 2.38 & 3.98  &  2.09 &  2.94 &   3.78 &  3.19   &   2.65  &  1.16 &  1.42 &   2.44  &    1.51   &     5.47   &    3.94  \\
\bottomrule
\end{tabular}
}
\vspace{-2mm}
\caption{Percentages of queries exhibiting different number of triples
  (in colors) for each dataset (top), the average numbers of triples
  of the queries for each dataset (\textit{Avg}$_{\#T}$, bottom), and
  the percentage of \select/\ask-queries (\textit{S/A},
  bottom).}\label{fig:triplecount}
\vspace{-4mm}
\end{figure*}

\subsection{Number of Triples in Queries}

In order to measure the size of the queries belonging to the datasets
under study, we have counted the total number of triples of the kind
$\langle s, p, o\rangle$ contained in \select and \ask queries.  In
this experiment, we merely counted the number of triples contained in
each query without further investigating the possible relationships
among them (such as join conditions, unions etc.), which are under
scrutiny later in the paper. We focus solely on \select and \ask
queries because these are the type of \sparql statements that truly
query the data, as opposed to \describe statements (which are
exploratory) and \construct statements (which construct
data).\footnote{For instance, 97\% of the \describe statements in our
  corpus do not have a body and therefore no triples.}

The plot in the upper part of Figure~\ref{fig:triplecount} illustrates
the results in terms of the percentages of \select and \ask queries (per dataset)
containing respectively from $0$ triples to a number of triples
greater than $11$.
A first observation that we can draw
from Figure~\ref{fig:triplecount} is that for the majority of the
datasets, the queries with a low number of triples (from $0$ to $2$)
have a noticeable share within the total amount of queries per
dataset.  Whereas these queries are almost the only queries present in
the \biopThir and \biopFour datasets, they have the least
concentration in \rkbeFour and \wikiSeven.
Both datasets have in fact unique
characteristics, \rkbeFour being a collection of queries with fixed
templates and \wikiSeven being the most diverse dataset of all,
gathering queries of rather disparate nature that are representatives
of classes of real queries issued on Wikidata.  Finally,
\dbpediaTwel--\dbpediaSixt, along with \lgdFour and \biomedThir are the
datasets exhibiting the most complex queries with extremely high numbers of
triples exceeding $11$.

We should note that \biomedThir has almost 85\% \describe queries and
2.42\% \construct queries. The numbers reported here only describe the
remaining 12.87\%. The table at the bottom of
Figure~\ref{fig:triplecount} shows the relative amounts of \select-
and \ask-queries per data set. It also shows the average number of
triples measured across all queries within each dataset. We can notice
a relative increase of this average for DPpedia from year $2014$ up to
year $2016$ and BioPortal and LGD in between years $2013$ and
$2014$. As expected, BioMed, BritishM and Wikidata have also
relatively high average number of triples, compared to the other
datasets for the reasons previously exposed.

Overall, we see that 56.45\% of the \select and \ask-queries in our
corpus use at most one triple, 90.76\% uses at most six triples, and
99.32\% at most twelve triples. The largest queries we found came from
\dbpediaFift (209 and 211 triples) and \biomedThir (221 and 229 triples).

\subsection{Operator Distribution}

\begin{table}[tb]
  \centering
  \begin{tabular}{@{}r <{\kern\tabcolsep} @{} R r@{}}
    \toprule
    \emph{Operator Set} & \emph{Absolute} & \emph{Relative}\\
    \midrule
    \emph{none}                                    &  17,482,313 &  33.49\%\\
    \textsf{F}                                     &   9,936,557 &  19.04\%\\
    \textsf{A}                                     &   3,911,748 &   7.49\%\\
    \textsf{A}, \textsf{F}                         &   3,261,138 &   6.25\%\\
    \rowcolor{black!10}[0pt][0pt] CPF subtotal     &  31,330,554 &  66.27\%\\
    \midrule
    \textsf{O}                                     &     542,900 &   1.04\%\\
    \textsf{O}, \textsf{F}                         &   1,791,512 &   3.43\%\\
    \textsf{A}, \textsf{O}                         &   1,728,907 &   3.31\%\\
    \textsf{A}, \textsf{O}, \textsf{F}             &     406,131 &   0.78\%\\
    \rowcolor{black!10}[0pt][0pt] CPF+\textsf{O}   &  +4,469,450 &  +8.56\%\\
    \midrule
    \textsf{G}                                     &   1,380,764 &   2.65\%\\
    \rowcolor{black!10}[0pt][0pt] CPF+\textsf{G}   &  +1,432,090 &  +2.74\%\\
    \midrule
    \textsf{U}                                     &   3,895,524 &   7.46\%\\
    \textsf{U}, \textsf{F}                         &     198,693 &   0.38\%\\
    \textsf{A}, \textsf{U}                         &     817,958 &   1.57\%\\
    \textsf{A}, \textsf{U}, \textsf{F}             &     812,381 &   1.56\%\\
    \rowcolor{black!10}[0pt][0pt] CPF+\textsf{U}   &  +5,724,556 & +10.97\%\\
    \midrule
    \textsf{A}, \textsf{O}, \textsf{U}, \textsf{F} &   4,084,154 &   7.82\%\\
  \end{tabular}
  \vspace{-2mm}
  \caption{Sets of operators used in queries: \filter (\textsf{F}), \and (\textsf{A}), \opt
    (\textsf{O}), \graph (\textsf{G}), and \union (\textsf{U})\label{tab:operators}}
  \vspace{-4mm}
\end{table}

In Table~\ref{tab:keywords} we see that \filter, \and, \union, \opt,
and \graph are used fairly commonly in the bodies of \select- and \ask
queries. We then investigated how these operators occur together. In
particular, we investigated for which queries the body \emph{only}
uses constructs with these operators.\footnote{There is one exception:
  For Wikidata, we removed SERVICE subqueries before the analysis
  (which appears in 222 of its queries and is used to change the
  language of the output).}  \footnote{This study closely follows a
  similar one \cite{PicalausaV-swim11} that was done on a log from
  DBpedia 2010. Our numbers should be compared to the numbers of ULog
  (the duplicate-free log) in \cite{PicalausaV-swim11}.}

The results are in Table~\ref{tab:operators}, which has two kinds
of rows. Each white row has, on its left, a set $S$ of operators from
$\cO = \{\filter, \and, \opt, \graph, \union\}$ and, on its right, the amount
of queries in our logs for which the body uses exactly the operators in $S$ (and none
from $\cO \setminus S$). The value for \emph{none} is the amount of
queries that do not use any of the operators in $\cO$ (including
queries that do not have a body).

Conjunctive patterns with filters are considered to be an important
fragment of SPARQL patterns, because they are believed to appear often
in practice~\cite{NeumannW-vldbj10,VidalRLMSP-eswc10}
\begin{definition}\upshape
  A \emph{conjunctive pattern with filters (\cpf)} is a graph pattern that
  only uses triples and the operators \and and \filter.
\end{definition}
Our logs contain
66.27\% \cpf patterns. Adding \opt to the \cpf fragment
would increase its relative size with 8.56\%, resulting in 74.83\% of
our queries. (Similarly for \graph and \union.)

Table~\ref{tab:operators} classifies
96.37\% of the \select- and \ask queries in our
corpus. The remaining queries either use other combinations from $\cO$
(0.30\%), use other features than those in $\cO$ in their body
(3.33\%) like \bind, \minus, subqueries or property paths.

There is a close relationship between \cpf patterns and
\emph{conjunctive queries} that, in some cases, can be extended to also
include queries with \opt and \graph. We discuss this in more detail
 in Section~\ref{sec:structural}.

\subsection{Subqueries and Projection}

Only 304,234 (0.54\%) queries in our corpus use subqueries. The
feature was most used in WikiData (9.74\%),
about an order of
magnitude more than in any of the other data sets.

Projection plays a crucial role in the complexity of
query evaluation. Many papers
\cite{BarceloPS-pods15,LetelierPPS-tods13,KaminskiK-icdt16,PerezAG-tods09,PicalausaV-swim11}
define evaluation as the following question: \emph{Given an RDF graph
  $G$, a graph pattern $P$, and a mapping $\mu$, is $\mu$ an answer to
  $P$ when evaluated on $G$?} In other words, the question is to
verify if a candidate answer $\mu$ is indeed an answer to the
query. If $P$ is a \cq, this problem is
\np-complete if the queries use projection
\cite{ChandraM-stoc77,BarceloPS-pods15,LeteliePPS-tods13}, but its complexity
drops to \ptime if projection is absent
\cite{PerezAG-tods09,BarceloPS-pods15,LetelierPPS-tods13}.\footnote{This
  difference can be understood as follows: If the query tests the
  presence of a $k$-clique, then without projection we are given a
  $k$-tuple of nodes and need to verify if they form a
  $k$-clique. With projection, we need to solve the NP-complete
  $k$-clique problem.} Therefore, the use of projection has a huge
influence of the complexity of query evaluation.

Surprisingly, we discovered that at least 14.98\% of the queries use
projection, which is about three times more than what Picalausa and
Vansummeren discovered in DBpedia logs from 2010
\cite{PicalausaV-swim11}.
The 14.98\% consists of 13.12\% \select queries plus
1.86\% \ask queries. Notice that the total number of \ask queries
(4.97\%) is significantly higher, even though they just return a
Boolean value and one would intuitively expect that almost all of them
would use projection. The reason is that most \ask queries do not use
variables: they ask if a concrete RDF triple is present in the
data. Following the test for projection in Section 18.2.1 in the
SPARQL recommendation \cite{sparql11}, we classified these queries as
not using projection.

Due to the use of the \bind operator, there was a number of queries
(1.3\%) where we could not determine if they use projection or
not. Therefore the number of queries with projection lies between
14.98\% and 16.28\%.

 \section{Structural Analysis}\label{sec:structural}

\sparql patterns of \select or \ask queries using only triple patterns
and the operators \and, \opt, and \filter (and, in particular, not
using subqueries or property paths) received considerable attention in
the literature (see, e.g.,
\cite{PerezAG-tods09,KaminskiK-icdt16,BarceloPS-pods15,KrollPS-icdt16,LetelierPPS-tods13}).
We refer to such patterns as \emph{\and/\opt/\filter patterns} or, for
succinctness, \emph{\aof patterns}.   Our corpus has 39.061.206
\aof patterns (74.83\% of the \select- and \ask queries).

In Section~\ref{sec:shape} we investigate the graph- and hypergraph
structure of \aof patterns. The graph structure gives us a clear view
on how such queries are structured and can tell us how complex such
queries are to evaluate. For a significant portion of queries,
however, the graph structure is not meaningful to capture their
complexity (cf.\ Example~\ref{ex:hypergraph}) and we therefore need to
turn to their hypergraph structure. Since the graph structure may be
easier to understand, we use the graph structure whenever we can.

We provide some background on the relationship between
the (hyper)graph structure of queries and the complexity of their evaluation.
Evaluation of \cqs is \np-complete in general
\cite{ChandraM-stoc77}, but becomes \ptime if their \emph{hypertree
  width} is bounded by a constant \cite{GottlobLS-jcss02}. Here, the
hypertree width measures how close the query is to a tree (the lower
the width, the closer the query is to a tree). Several
state-of-the-art join evaluation algorithms (e.g.,
\cite{AbergerTOR-sigmod16,KalinskyEK-edbt17}) effectively use the
hypergraph structure of queries to improve their performance, even in
the context of RDF processing \cite{AbergerTOR-icdew16}. We establish
in Section~\ref{sec:gmark} that there are significant performance
differences in today's query engines, even when the hypertreewidth of
queries just increases from one to two.

\paragraph*{Graph and Hypergraph of a Query}
We first make more precise what we mean by the graph and hypergraph of
a query. An \emph{(undirected) graph} $G$
is a pair
$(V,E)$ where $V$ is its (finite) set of nodes and $E$ is its set of
edges, where an edge $e$ is a set of one or two nodes, i.e., $e \subseteq V$
and $|e| = 1$ or $|e| = 2$. A \emph{hypergraph} $\cH$ consists of a
(finite) set of nodes $\cV$ and a set of hyperedges $\cE \subseteq 2^\cV$,
that is, a hyperedge is a set of nodes.

Most \sparql patterns do not use variables as predicates, that is, they use
triple patterns $(s,p,o)$ where $p$ is an IRI. We call such patterns
\emph{graph patterns}. Evaluation of graph patterns is tightly
connected to finding embeddings of the graph representation of the
query into the data.\footnote{In particular, it consists of finding
  embeddings of the directed and edge-labeled variant of the graph,
  but we omit the edge directions and -labels for simplicity. They do not
  influence the structure and cyclicity of graph patterns.} We define the \emph{canonical graph} of graph pattern
  $P$ to be the following graph: $E =
\{\{x,y\}) \mid \ell$ is a literal and $(x, \ell, y)$
is a triple pattern in $P\}$ and $V = \{x \mid (x,\ell,y) \in E$ or
$(y,\ell,x) \in E\}$.

Hypergraph representations can be considered for all \aof patterns.
The \emph{canonical hypergraph of a pattern $P$} is defined as $\cE = \{X \mid
X$ is the set of blank nodes and variables appearing in a triple
pattern in $P\}$ and $\cV = \cup_{e \in E} e$.

\begin{figure}
  \centering
  \begin{tikzpicture}
    \node (g1) at (0,1) {$x1$};
    \node [right of=g1] (g2)          {$x2$};
    \node [right of=g2] (g3)          {$x3$};
    \node [right of=g3] (g4)          {$x4$};

    \path[-] (g1) edge node [above] {:$a$} (g2)
              (g2) edge node [above] {:$b$} (g3)
              (g3) edge node [above] {:$c$} (g4)
              ;

    \node              (1) at (0,0) {$x1$};
    \node [right of=1] (2)          {$x3$};
    \node [right of=2] (3)          {$x4$};
    \node [right of=3] (4)          {$x5$};

    \path[-] (1) edge node [above] {$x2$} (2)
              (2) edge node [above] {:$a$} (3)
              (3) edge node [above] {$x2$} (4)
              ;
  \end{tikzpicture}
  \qquad
  \begin{tikzpicture}
    \node              (v) at (0,0) {$x2$};
    \node [right of=v] (u)          {$x1$};
    \node [right of=u] (w)          {$x3$};
    \node [below of=u,node distance=1.6cm] (y) {$x4$};
    \node at ($(w)!.5!(y)$) (x)     {};
    \node at ($(y)!.5!(v)$) (z)     {$x5$};

    \node[draw,shape=rounded rectangle, inner xsep=13mm, inner ysep=3mm] at (u) {};
    \node[draw,shape=rounded rectangle, rotate=60, inner xsep=13mm, inner ysep=3mm] at (x) {};
    \node[draw,shape=rounded rectangle, rotate=-60, inner xsep=13mm, inner ysep=3mm] at (z) {};

\end{tikzpicture}
  \caption{Canonical graphs and hypergraph for queries in Example~\ref{ex:hypergraph}.\label{fig:hypergraph}}
\end{figure}
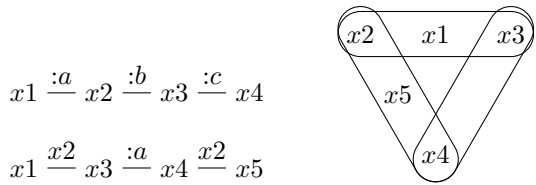

\begin{example}\label{ex:hypergraph}
Consider the following (synthetic) queries:
\begin{verbatim}
ASK WHERE {?x1 :a ?x2 . ?x2 :b ?x3 . ?x3 :c ?x4}
ASK WHERE {?x1 ?x2 ?x3 . ?x3 :a ?x4 . ?x4 ?x2 ?x5}
\end{verbatim}
Figure~\ref{fig:hypergraph} (top left) depicts the canonical graph of
the first query, which is a sequence of three edges. (We annotated the
edges with their labels in the query to improve understanding.) The
bottom left graph in Figure~\ref{fig:hypergraph} shows why we do not
consider canonical graphs for queries with variables on the predicate
position in triples. The topological structure of this graph is, just
as for the first query, a sequence of three edges, which completely
ignores the join condition on $?x2$. For this
query, the canonical hypergraph in Figure 2 (right) correctly captures
the cyclicity of the query.
\end{example}

\subsection{Comparative Evaluation of Chain and Cycle Queries}\label{sec:gmark}
We conducted a set of experiments aiming at comparing the execution
times of conjunctive queries whose their corresponding canonical graphs
exhibit specific shapes. We have chosen chain and cycle queries in this empirical study.  A
\emph{chain query (of length $k$)} is a \cq for which the canonical graph is
isomorphic to the undirected graph with edges
$\{x_0,x_1\},\{x_1,x_2\},\ldots,\{x_{k-1},x_k\}$. (The first query in
Example~\ref{ex:hypergraph} is a chain query of length three.) A
\emph{cycle query (of length $k$)} is a \cq for which the canonical
graph is isomorphic to $\{x_0,x_1\},\ldots,\{x_{k-1},x_0\}$.
These shapes have been selected as representatives of the queries with hypertreewidth
$1$ and $2$, respectively, and have also been used to compare the performances of join algorithms in
other studies, e.g., \cite{KalinskyEK-edbt17}. In order to generate
query workloads containing the aforementioned types of queries, we
have used gMark~\cite{BaganBCFLA17}, a publicly
available\footnote{\url{https://github.com/graphMark/gmark}}
schema-driven generator for graph instances and graph queries. We
tuned gMark to generate diverse query workloads, each containing $100$ chain
and cycle queries, respectively.\footnote{We recall that gMark can generate queries of four
  shapes: chain, star, chain-star and cycle. We have thus
  cherry-picked chain queries as representatives of queries with
  hypertreewidth equal to $1$.} Each workload has been generated by using chains and cycles
  of different length varying from $3$ to $8$.  In these experiments, we have
considered and contrasted two opposite graph database systems, namely
PostgreSQL~\cite{postgres}, an open-source relational DBMS, and
BlazeGraph~\cite{blazegraph}, an high-performance SPARQL query engine
powering the Wikimedia's official query service~\cite{VrandecicK14}
and thus used for Wikidata real-world queries.  We have run these
experiments on 2-CPUs Intel Xeon E5-2630v2 2.6 GHz
server\footnote{Every CPU has 6 physical cores and (with
  hyperthreading) 12 logical cores.} with 128GB RAM and running Ubuntu
16.04 LTS.  We used PostgreSQL v.9.3 and Blazegraph v.2.1.4
for the experimental setup.
We employed the Bib use case in the gMark configuration~\cite{BaganBCFLA17} for the schema of the generated
graph (of size $100k$ nodes) and of the generated queries as well.
We employed
the query workloads in SQL and SPARQL as generated by gMark after
elimination of empty unions (since gMark is geared towards generating
UCRPQs) and of the keyword \distinct in the body of the queries.  Since
gMark allowed us to obtain mixed workloads of \select/\ask queries and
we wanted to focus on one query type at a time, we manually replaced
the \select clauses with compatible \ask clauses (and, vice versa for
full workloads of \select queries, whose results are comparable and omitted for space reasons).
Figure~\ref{fig:gmark:final} (top) depicts the average runtime (in ns, logscale) of
our workloads of chain (cycle, resp.) queries with length from $3$ to $8$ on
Blazegraph (BG) and PostgreSQL (PG). We can observe that the overall
performance of BG is superior to that of PG. Indeed, in PG many cycles queries are
timed out (after 300s per query) and we expect that the real overall performance of PG is
even worse than the results reported in Figure~\ref{fig:gmark:final}.
Figure~\ref{fig:gmark:final} (bottom) reports the reached timeouts for workloads of cycle queries of various sizes when executed in PG.
It is worthwhile observing that for both systems the difference between average runtime of
chain query workloads and cycle query workloads is non negligible, thus confirming that
we cannot ignore the graph representation and the shape of queries.
This experiment also motivated us to dig deeper in the shape analysis of our query logs,
which we report in Section~\ref{sec:shape}.

\pgfplotsset{filter discard warning=false}

\newcommand{\filtertableplot}[3]{
\pgfplotstablegetelem{\coordindex}{#2}\of{#1}
\IfStrEq{\pgfplotsretval}{#3}
{}
{\def\pgfmathresult{}}
}

\begin{figure}
\begin{tikzpicture}
	\begin{axis}[
	width=\linewidth,
	height=7cm,
        ymode=log,
	ylabel={Avg. query runtime per workload W-x (in ns)},
		legend style={at={(0.5,-0.15)},
		anchor=north,legend columns=-1},
		xticklabels={
          ,,\small{W-3},
          \small{W-4},
          \small{W-5},
          \small{W-6},
          \small{W-7},
          \small{W-8},,
        },
        xtick={1, ..., 8},
        mark size=3.0pt,
	scatter/classes={
		a={mark=square*,blue},
		b={mark=diamond*,red},
		c={mark=square*,draw=black,fill opacity=0.5},
		d={mark=diamond*,brown!60!black}
		}]
	\addplot[scatter,only marks,
		scatter src=explicit symbolic]
	table[meta=label] {bench_chainBG.tsv};
	\addlegendentry{chainBG}
	\addplot[scatter,only marks,
		scatter src=explicit symbolic]
	table[meta=label] {bench_chainPG.tsv};
	\addlegendentry{chainPG}
	\addplot[scatter,only marks, fill opacity = 0.5,
		scatter src=explicit symbolic]
	table[meta=label] {bench_cycleBG.tsv};
	\addlegendentry{cycleBG}
	\addplot[scatter,only marks,
		scatter src=explicit symbolic]
	table[meta=label] {bench_cyclePG.tsv};
	\addlegendentry{cyclePG*}
	\end{axis}
\end{tikzpicture}

\centering
\begin{tabular}{@{}ccccccc@{}}
W-x &  W-3 & W-4 &  W-5 &  W-6 &  W-7 & W-8 \\
\hline
\%t/o & 18\% & 34\% & 43\% & 39\% & 43\% & 30\%
\end{tabular}

\caption{Execution times (top) of diverse workload of chain/cycle
queries (of length 3,4,5,6) on Blazegraph (BG) and Postgresql
(PG). Number of timeouts per
workload for CyclePG only (bottom). CyclePG times include t/o of 300s (per query).}\label{fig:gmark:final}
\end{figure}
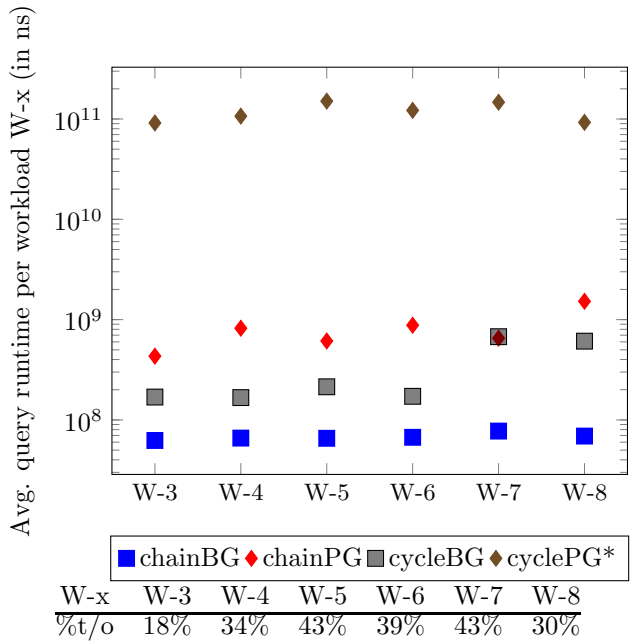

\subsection{Classes of Queries for (Hyper)graphs}

We now discuss the classes of queries for which we will investigate
graph- and hypergraph structures in Section~\ref{sec:shape}. To the
best of our knowledge, all the literature relating (hyper)graph structure of
queries to efficient evaluation was done on \aof patterns, which is why we
only consider \aof patterns here.  The simplest such queries are the
\cqs, which motivated the classical literature on query evaluation and
hypertree structure \cite{ChandraM-stoc77,GottlobLS-jcss02}.  We
discovered that
54.58\% of the \aof patterns are \cqs.

Next, we extend \cqs with \filter and \opt such that the relationship
between efficient query evaluation and their (hyper)graph structure is still similar as for
\cqs. However, this requires some care, especially when considering
\opt~\cite{BarceloPS-pods15,PerezAG-tods09}.

We first define a fragment of CPF patterns that can be readily
translated to \cqs and can be evaluated similarly. We say that a
filter constraint $R$ is \emph{simple} if vars$(R)$ contains at most
one variable or is of the form $?x = ?y$.\footnote{If we encounter a
  filter constraint of the form $?x = ?y$, we collapse the nodes $?x$
  and $?y$ in the graph and hypergraph of the query.} (An almost
identical class of queries was also considered in \cite{PicalausaV-swim11}.)
\begin{definition}\upshape
  A \emph{conjunctive query with filters (\cqf query)} is a CPF pattern that
  only uses simple filters.
\end{definition}
In our corpus,
84.08\% of the \aof patterns are in \cqf.

We now additionally consider \opt. P\'erez et
al.~\cite{PerezAG-tods09} showed that unrestricted use of \opt in
graph patterns makes query evaluation \pspace-complete, which is
significantly more complex than the \np-completeness of \cqf queries. They discovered that
patterns that satisfy an extra condition called
\emph{well-designedness} \cite{PerezAG-tods09}, can be evaluated much
more efficiently. Letelier et al.\ show that, in the presence of projection,
evaluation of well-designed patterns is $\sigmatwo$-complete
\cite{LetelierPPS-tods13}.
\begin{definition}\upshape
  A graph pattern $P$ using only the operators \and, \filter, and \opt
  is \emph{well-de\-signed} if for
  every occurrence $i$ of an \opt-pattern $(P_1$ \opt $P_2)$ in $P$, the
  variables from $\vars(P_2) \setminus \vars(P_1)$ occur in $P$ only
  inside $i$.\footnote{Perez et al.'s definition also has a safety condition
    on the filter statements of the patterns, but the omission of this
  condition does not affect the results in this paper.}
\end{definition}
In our corpus, 98.53\% of the \aof patterns are well-designed
(but do not necessarily have simple filters).
Unfortunately, it is not yet sufficient for well-designed patterns to
have a hypergraph of constant hypertreewidth for their evaluation to
be tractable \cite{BarceloPS-pods15}. However, Barcel\'o et al.\ show
that this can be mended by an additional restriction called \emph{bounded
  interface width}. We explain this notion by example and refer to \cite{BarceloPS-pods15} for
details.

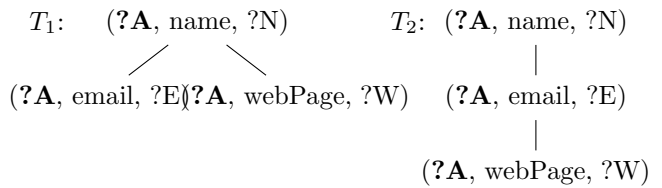
\begin{figure}
    \begin{tikzpicture}[level distance=10mm
      ]
      \node {(\textbf{?A}, name, ?N)}
      child [sibling distance=26mm] { node {(\textbf{?A}, email, ?E)}}
      child [sibling distance=26mm] { node {(\textbf{?A}, webPage, ?W)}}
      ;

      \node at (4.5,0) {(\textbf{?A}, name, ?N)}
      child { node {(\textbf{?A}, email, ?E)}
        child { node {(\textbf{?A}, webPage, ?W)}}
      }
      ;

      \node at (-2,0) {$T_1$:};
      \node at (2.8,0) {$T_2$:};
   \end{tikzpicture}
   \caption{Pattern trees that correspond to the queries in
     Example~\ref{ex:queries-for-trees}}\label{fig:pattern-trees}
\end{figure}

\begin{example}\label{ex:queries-for-trees}\upshape
  The following patterns come from \cite{PerezAG-tods09,LetelierPPS-tods13}:

  \noindent $P_1 = (((?A,$ name, $?N)$ \opt $(?A,$ email, $?E))$

  \hfill \opt $(?A,$
  webPage, $?W))$

  \noindent and $P_2 = ((?A,$ name, $?N)$

  \hfill \opt $((?A,$ email, $?E)$ \opt $(?A,$ webPage, $?W)))$

  \noindent Figure~\ref{fig:pattern-trees} has tree representations
  $T_1$ and $T_2$ for $P_1$ and $P_2$, respectively, called
  \emph{pattern trees}. The pattern trees $T_i$ are obtained from the parse trees of $P_i$ by
  applying a standard encoding based on Currying~\cite[Section
  4.1.1]{MartensN-jcss07}.  The encoding only affects the arguments of
  the \opt operators in the queries. If the query also uses \and, then
  it should first be brought in \emph{\opt-normal
    form}~\cite{PerezAG-tods09} and then turned into a pattern
  tree. The resulting pattern trees will then have a
  \cq in each of its nodes.

  Barcel\'o et al.\ define pattern trees to be \emph{well-designed}
  if, for each variable, the set of nodes in which
  it occurs forms a connected set. Notice that this is the case for
  $T_1$ and $T_2$. It would be violated in $T_1$ if the root would not
  use the variable $?A$. Likewise, it would be violated in $T_2$ if the node labeled $(?A$, email,
  $?E)$ would not use the variable $?A$.

  The \emph{interface width} of the pattern trees is the maximum
  number of common variables between a node and its child.
  Both trees
  in Figure~\ref{fig:pattern-trees} (and both queries $P_1$ and $P_2$)
  therefore have interface width one. (Common variables are bold in Figure~\ref{fig:pattern-trees}.)
 If $T_1$ would use variable $?W$
  instead of $?N$, then its interface width would be two.
\end{example}

\begin{definition}\upshape
  A graph pattern $P$ using only the operators \and, \filter, and \opt
  is in \cqof if it has a well-designed pattern tree with interface width 1.
\end{definition}
Perhaps surprisingly, out of all queries that are well-designed and
have simple filters, we only found 310 queries that had an interface
width more than one.
In fact, 93.87\% of the \aof patterns are \cqof queries.

 \section{Shape Analysis}\label{sec:shape}

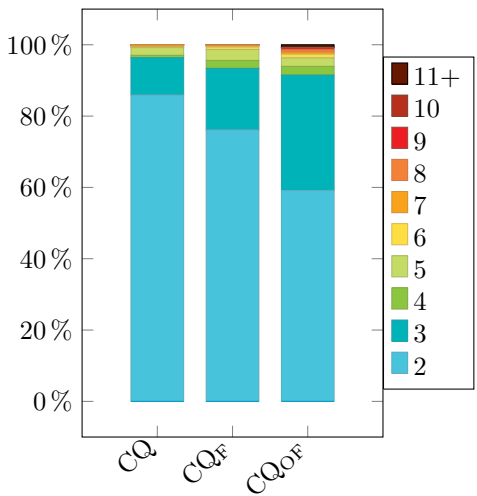
\begin{figure}
  \centering
  \pgfplotstableread[col sep=comma]
  {triples-nontriv-cqs.csv}
  {\datatableE}

  \begin{tikzpicture}
    \begin{axis}[
      enlarge x limits=0.5,
      ybar stacked,
      legend style={at={(1,0.5)}, anchor=west},
      reverse legend,
      ylabel near ticks,
      yticklabel=\pgfmathparse{100*\tick}\pgfmathprintnumber{\pgfmathresult}\,\%,
      bar width=0.7cm,
      x=1cm,
      xticklabels={
        \cq,
        \cqf,
        \cqof
      },
      xtick={1, ..., 3},
      x tick label style={rotate=45,anchor=east},
      legend cell align=left
      ]
      \addplot+[ProcessBlue!80!black,fill=ProcessBlue,ybar] table[x=id, y=1] from {\datatableE};
      \addplot+[SkyBlue!80!black,fill=SkyBlue,ybar] table[x=id, y=2] from {\datatableE};
      \addplot+[BlueGreen!80!black,fill=BlueGreen,ybar] table[x=id, y=3] from {\datatableE};
      \addplot+[LimeGreen!80!black,fill=LimeGreen,ybar] table[x=id, y=4] from {\datatableE};
      \addplot+[SpringGreen!80!black,fill=SpringGreen] table[x=id, y=5] from {\datatableE};
      \addplot+[Goldenrod!80!black,fill=Goldenrod] table[x=id, y=6] from {\datatableE};
      \addplot+[YellowOrange!80!black,fill=YellowOrange,ybar] table[x=id, y=7] from {\datatableE};
      \addplot+[Orange!80!black,fill=Orange,ybar] table[x=id, y=8] from {\datatableE};
      \addplot+[Red!50!black,fill=Red,ybar] table[x=id, y=9] from {\datatableE};
      \addplot+[BrickRed!50!black,fill=BrickRed,ybar] table[x=id, y=10] from {\datatableE};
      \addplot+[black,fill=Sepia,ybar] table[x=id, y=11] from {\datatableE};

      \legend{2, 3, 4, 5, 6, 7, 8, 9, 10, $11+$}
    \end{axis}
\end{tikzpicture}

  \caption{Size of \cq-like queries with at least two
    triples. \label{fig:cq-size}}
\end{figure}

In this section we analyze the shapes of the canonical graphs and the
tree- and hypertree width of \cqs, \cqf queries, and \cqof queries.
We start with a note on the size of these
queries. Figure~\ref{fig:cq-size} shows the respective sizes of these
queries that have at least two triples. The fractions of queries with
one triple are 82\%, 83.45\%, and 75.52\%  for \cq, \cqf, and \cqof,
respectively. Unsurprisingly, small queries are more likely to be in
one of these fragments and, therefore, simple queries are represented
even more in these data sets than in the overall data
set. Nevertheless, we have \cqs and \cqf queries with up to 81 triples
and \cqof queries with up to 229 triples.

\subsection{Graph Structure}\label{sec:graphstructure}

We analyse the graph structure of queries. Recall that we only
consider graph shapes for queries that do not use variables in the
predicate position of triples, for reasons explained in
Section~\ref{sec:structural}. We consider the remaining 6.96 million
queries in \cqof in Section~\ref{sec:hypergraph}.

We first recall or define the basic shapes of the canonical graphs
that we will study in this section.  The shapes \emph{chains} and
\emph{cycle} are already defined in Section~\ref{sec:gmark}. A
\emph{chain set} is a graph in which every connected component is a
chain. (So, each chain is also a chain set.)

A \emph{tree} is an undirected graph such that, for every pair of
nodes $x$ and $y$, there exists exactly one undirected path from $x$
to $y$. A \emph{forest} is a graph in which every connected component
is a tree.

A \emph{star} is a tree for which there exists exactly one node with
more than two neighbors, that is, there is exactly one node $u$ such
that there exist $u_1$, $u_2$, and $u_3$, all pairwise different and
different from $u$, for which $\{u,u_i\} \in E$ for each $i = 1,2,3$.

Inspired by the results obtained with gMark on synthetic queries, we
proceeded with the analysis of the query logs by looking at the
encountered query shapes. Here, we consider queries as
edge-labeled graphs, as defined in Section~\ref{sec:structural}. In
the next subsection we also investigate the hypergraph structure.

We investigate \cqs, \cqf queries, and \cqof queries.
The last two fragments are interesting in that they bring under scrutiny more
queries than the plain \cq set of query logs (by an increase of roughly
40\% and 47\%, respectively).
We first wanted to identify classical query
shapes, such as all variants of tree-like shapes (single edges, chains,
sets of chains, stars, trees, and forests). The results are summarized
in the three tables in Table~\ref{tab:shape}.
From the analysis, we can draw
the following observations. While tree-shaped queries even in their simple
forms (chain of length 1 or single edges) are very frequent, the only
observed exception occurs with star queries,
which have very low
occurrence with respect to the other tree-like shapes.

Since simple queries are overrepresented in query logs (already over
80\% of \cqf patterns uses only one triple, for example), it is no
surprise that the overwhelming majority of the queries is acyclic,
i.e., a forest. However, we also wanted to get a better understanding
of the more \emph{complex} queries in the logs, so we also
investigated the cyclic queries. Our goal is to obtain a cumulative
shape analysis where simpler shapes are subsumed by more sophisticated
query shapes, with the latter reaching almost 100\% coverage of the
query logs.

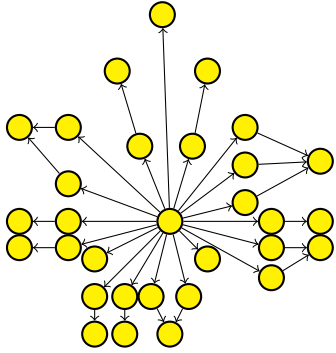
\begin{figure}[t]
\centering
\begin{tikzpicture}
        \tikzstyle{sommet}=[circle,draw,thick,fill=yellow]

	\node[sommet] (v1) at (0.3,0) {};
	\node[sommet] (v2) at (0.9, 0.75) {};
	\node[sommet] (v3) at (1.5,0) {};
	\node[sommet] (v4) at (0.6,-1) {};
	\node[sommet] (v5) at (1,-2) {};
	\node[sommet] (v6) at (1.3,-1) {};
	\node[sommet] (v7) at (-0.35,-0.75) {};
	\node[sommet] (v8) at (-1,-0.75) {};
	\node[sommet] (v9) at (-0.35,-1.5) {};
	\node[sommet] (v10) at (2,-0.75) {};
	\node[sommet] (v11) at (2,-1.25) {};
	\node[sommet] (v12) at (2,-1.75) {};
	\node[sommet] (v13) at (3,-1.2) {};
	\node[sommet] (v14) at (-0.35,-2) {};
	\node[sommet] (v15) at (-1,-2) {};
	\node[sommet] (v16) at (2.35,-2) {};
	\node[sommet] (v17) at (3,-2) {};
	\node[sommet] (v18) at (2.35,-2.35) {};
	\node[sommet] (v19) at (3,-2.35) {};
	\node[sommet] (v20) at (2.35,-2.75) {};
	\node[sommet] (v21) at (1.5,-2.5) {};
	\node[sommet] (v22) at (1.25,-3) {};
	\node[sommet] (v23) at (1,-3.5) {};
	\node[sommet] (v24) at (0.75,-3) {};
	\node[sommet] (v25) at (-0.35,-2.35) {};
	\node[sommet] (v26) at (-1,-2.35) {};
         \node[sommet] (v27) at (0,-2.5) {};
         \node[sommet] (v28) at (0.4,-3) {};
	\node[sommet] (v29) at (0,-3) {};
	\node[sommet] (v30) at (0.4,-3.5) {};
	\node[sommet] (v31) at (0,-3.5) {};
	\path[->]
(v4) edge [above] node [align=center]  {} (v1)
(v5) edge [above] node [align=center]  {} (v4)
(v5) edge [above] node [align=center]  {} (v2)
(v5) edge [above] node [align=center]  {} (v6)
(v6) edge [above] node [align=center]  {} (v3)
(v5) edge [above] node [align=center]  {} (v7)
(v5) edge [above] node [align=center]  {} (v9)
(v7) edge [above] node [align=center]  {} (v8)
(v9) edge [above] node [align=center]  {} (v8)
(v5) edge [above] node [align=center]  {} (v10)
(v5) edge [above] node [align=center]  {} (v11)
(v5) edge [above] node [align=center]  {} (v12)
(v10) edge [left] node [align=center]  {} (v13)
(v11) edge [left] node [align=center]  {} (v13)
(v12) edge [left] node [align=center]  {} (v13)
(v5) edge [above] node [align=center]  {} (v16)
(v16) edge [above] node [align=center]  {} (v17)
(v5) edge [above] node [align=center]  {} (v14)
(v14) edge [above] node [align=center]  {} (v15)
(v5) edge [above] node [align=center]  {} (v18)
(v5) edge [above] node [align=center]  {} (v20)
(v18) edge [above] node [align=center]  {} (v19)
(v20) edge [above] node [align=center]  {} (v19)
(v5) edge [above] node [align=center]  {} (v21)
(v5) edge [above] node [align=center]  {} (v22)
(v5) edge [above] node [align=center]  {} (v24)
(v22) edge [above] node [align=center]  {} (v23)
(v24) edge [above] node [align=center]  {} (v23)
(v5) edge [above] node [align=center]  {} (v25)
(v25) edge [above] node [align=center]  {} (v26)
(v5) edge [above] node [align=center]  {} (v27)
(v5) edge [above] node [align=center]  {} (v28)
(v5) edge [above] node [align=center]  {} (v29)
(v28) edge [above] node [align=center]  {} (v30)
(v29) edge [above] node [align=center]  {} (v31)
;
\end{tikzpicture}
\caption{\label{fig:example-flower}An example of flower query found in our DBPedia
query logs (we added arrows to indicate the edge directions in the query; labels are omitted for confidentiality reasons).
}
\end{figure}

A first observation was that plain cycles are not very common.
By visually inspecting the remaining cyclic queries, we observed that
many of them could be seen as a node with simple attachments, which we
call \emph{flower}.
\begin{definition}\upshape\label{def:flower}
  A \emph{petal} is a graph consisting of a source node $s$, target
  node $t$, and a set of at least two node-disjoint paths from $s$ to $t$. (For
  instance, a cycle is a petal that uses two paths.)
    A \emph{flower} is a graph consisting of a node $x$ with three types of
  attachments: chains (the \emph{stamens}), trees that are not chains (the \emph{stems}),
  and \emph{petals}.
\end{definition}
An example of a real flower query posed by users in one of our
DBpedia logs is illustrated in Figure~\ref{fig:example-flower}. It
consists of a central node with four petals (one of which using three paths), ten stamens and zero
stems attached.

We also considered sets of flowers, which we called \emph{flower
  sets}, to further increase the ratio of queries that could be
classified from the original logs. The number of flowers and flower
sets in the query logs overcome those of trees and forests by roughly
0.05\%, respectively for $CQ$, $CQ_{F}$ and $CQ_{OF}$, and for all the
three fragments flowerSets queries could get significantly closer to
100\% coverage than plain forests.

\begin{table*}[t]
  \centering
  \begin{tabular}{@{}r <{\kern\tabcolsep} @{} R r@{}}
    & \cq \\
    \toprule
    \emph{Shape} &   \emph{$\#$Queries} &  \emph{Relative $\%$}\\
    \midrule
    single edge & 12,273,871 &	77.98\%\\
    chain       &     15,561,944	& 98.87\%\\
    chain set    &   15,570,042	& 98.93\%\\
    star      &         147,457	&  0.94\%\\
    tree           &  15,723,163 &   99.90\%\\
    forest        &  15,731,535 & 99.95\%\\
    \midrule
    cycle        &      4,550	&  0.03\%\\
    flower         &     15,730,043 	& 99.94\%\\
    flower set  &  15,738,439	& 100.00\%\\
    treewidth $\leq 2$ &  15,739,056 & 100.00\%\\
    \midrule
    treewidth $= 3$ & 1 & 0.00\%\\
    \rowcolor{black!10}[0pt][0pt] total      &     15,739,057  &
    100.00\%\\
  \end{tabular}
  \quad
  \begin{tabular}{@{}r <{\kern\tabcolsep} @{} r@{}}
    \multicolumn{2}{c}{\cqf}\\
    \toprule
    \emph{$\#$Queries} &  \emph{Relative $\%$}\\
    \midrule
    21,198,951	& 81.04\%\\
    25,403,669 &	97.12\%\\
    25,418,689  &	 97.17\%\\
    702,228 &	2.68\%\\
    26,127,544	& 99.88\%\\
    26,143,128 &	99.94\%\\
    \midrule
    4,705 & 0.02\%\\
    26,135,676	& 99.92\%\\
    26,151,291	& 99.97\%\\
    26,157,879 & 100.00\%\\
    \midrule
     1 & 0.00\%\\
    \rowcolor{black!10}[0pt][0pt]       26,157,880	&   100.00\%\\
  \end{tabular}
  \quad
  \begin{tabular}{@{}r <{\kern\tabcolsep}@{} r@{}}
    \multicolumn{2}{c}{\cqof}\\
    \toprule
    \emph{$\#$Queries} &  \emph{Relative $\%$}\\
    \midrule
    21,479,706	& 72.30\%\\
    26,887,865 & 90.50\%\\
    26,937,578 &	 90.67\%\\
    2,654,497 &	8.94\%\\
    29,599,539 & 99.63\%\\
    29,651,600 &  99.81\%\\
    \midrule
    4,734 & 0.02\%\\
    29,614,330	& 99.68\%\\
    29,666,423  & 99.86\%\\
     29,708,967	&   100.00\%\\
     \midrule
      1 & 0.00\%\\
     \rowcolor{black!10}[0pt][0pt]    29,708,968	&   100.00\%\\
  \end{tabular}
\vspace{-0.18cm}
\caption{Cumulative shape analysis of $CQ$, $CQ_{F}$, $CQ_{FO}$ across
  all logs.
}
\label{tab:shape} \label{tab:shape-noConsts}
\end{table*}

In the above analysis, we have analyzed the shapes of queries when the
latter are represented as canonical graphs as defined in
Section~\ref{sec:structural}, i.e., the nodes can be
either variables or constants.
Constants are in fact necessary to fully characterize query shapes,
even though they do not play a major role in query optimization, as
variables do. For that reason, we have rerun the above analysis on
queries excluding constants in order to identify the differences in
the obtained shape classification. The most significant observation
here was that 9.66 million single edge \cqs (78.70\% of the single
edge \cqs) uses constants.

For the queries with cycles, we also investigate what is the length of
the \emph{shortest cycle} in the query. We discovered, for 39,471
queries, the shortest cycle has length three. For 6,561 and 5,733
queries, the shortest cycles had length 4 and 5, respectively. For 26
queries, the length was larger. We found two queries for which the
shortest cycle was 14, which is the largest value we found.

\subsection{Tree- and Hypertreewidth}\label{sec:hypergraph}
It is well-known that the tree- or hypertreewidth of queries are
important indicators to gauge the complexity of their evaluation. We
therefore investigated the tree- and hypertreewidth the \cqs, \cqf-
and \cqfo queries. We do not formally define tree- or hypertreewidth
in this paper but instead refer to an excellent introduction
\cite{DBLP:books/cu/p/GottlobGS14}. In the terminology of Gottlob et
al., we investigate the \emph{generalized hypertree width} of the
canonical hypergraphs of queries.

\paragraph*{Treewidth}
All shapes we discussed in Section~\ref{sec:graphstructure} have tree\-width at most
two. Forests (and all subclasses thereof) have tree\-width one, whereas
cycles, flowers, and flower sets have treewidth two. We investigated
the remaining queries by hand and discovered that one query had
treewidth three and all others had treewidth two, see
Table~\ref{tab:shape}. From the treewidth perspective, it is
interesting to note that many queries of treewidth two are
\emph{flowers} or \emph{flower sets}
(Definition~\ref{def:flower}), which are a very restricted fragment.

\paragraph*{Hypertree Width}
We recall that we only considered canonical graphs for queries that do
not use variables in the predicate position of triple patterns. In
\cqof, 6,959,510 queries used this feature and therefore
we must consider the hypergraph structure to correctly measure the
cyclicity of these queries. We determined their (generalized)
hypertree width with the tool \texttt{detkdecomp} from the Hypertree
Decompositions home page \cite{detkdecomp}.

Our results are as follows. All the remaining queries had
hypertree width one, except for 86 queries with hypertree width two
and eight queries with hypertree width three.

We also looked at the number of nodes in the hypertree decompositions
that the tool gave us, since this number can be a guide for how well
\emph{caching} can be exploited for query evaluation \cite{KalinskyEK-edbt17} (the
higher the number, the better caching can be exploited). For the
queries with hypertree width one, the number of nodes in the
decompositions corresponds to their number of edges, which can already
be seen in Figure~\ref{fig:cq-size}. (Nevertheless, we found several hundred
queries with more than 100 nodes in their hypertree decompositions,
all of them occurring in \dbpediaFift and \dbpediaSixt.) Finally, we
observed that the queries
with hypertree width two and three both had decompositions with up to
ten nodes, respectively.

\begin{figure}[t]
\centering
\begin{tikzpicture}
        \tikzstyle{sommet}=[circle,draw,thick,fill=yellow]

	\node[sommet][label={\scriptsize ?subject\_nationality}]  (v1)  at (3,.8) {};
	\node[sommet][label={\scriptsize ?subject\_birthPlace}](v2) at (-3,.8) {};
	\node[sommet][label={\scriptsize ?subject\_genre}](v3) at (0,.8) {};
	\node[sommet][label=below:{\scriptsize ?object\_genre}] (v4) at (3,-.8) {};
	\node[sommet][label=below:{\scriptsize ?object\_birthPlace}] (v5) at (-3,-.8) {};
	\node[sommet][label=below:{\scriptsize ?object\_nationality}] (v6) at (0,-.8) {};
	\path[->]
(v1) edge [above] node [align=center]  {} (v4)
(v1) edge [above] node [align=center]  {} (v5)
(v1) edge [above] node [align=center]  {} (v6)
(v2) edge [above] node [align=center]  {} (v4)
(v2) edge [above] node [align=center]  {} (v5)
(v2) edge [above] node [align=center]  {} (v6)
(v3) edge [above] node [align=center]  {} (v4)
(v3) edge [above] node [align=center]  {} (v5)
(v3) edge [above] node [align=center]  {} (v6)
;
\end{tikzpicture}
\caption{\label{fig:example-queryTW} The DBPedia
query exhibiting tree width equal to 3.
\label{fig:treewidth-three}}
\end{figure}
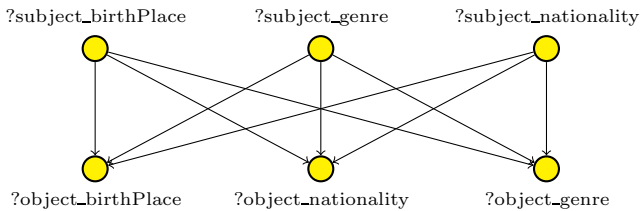

 \section{Property Paths}

We found 247,404 property paths in our corpus. Although property paths
are therefore rare in relation to the entire corpus, this is not so
for every data set: 92 queries
(29.87\%) in \wikiSeven have property paths.

A large fraction of these
property paths are extremely simple. For instance, 63,039 property
paths are $!a$ (``follow an edge not labeled $a$'') and 306 are
\^{}$a$ (``follow an $a$-edge in reverse direction''). In the
following, we focus on the remaining 184,059 property paths, which
express queries on the graph that do more than simply follow an edge
(such queries are sometimes called \emph{navigational queries}).

Here, 66,262 (36\%) use reverse navigation, i.e., the operator
``\^{}'', within more complex expressions.
In Table \ref{fig:propertypaths2}, we present an overview of the
property paths different from $!a$ and
\^{}$a$. In our classification, we treat \^{}$a$ and $!a$ the same as
a literal. For instance, we classify $a/b$, $($\^{}$a)/b$, and
$(!a)/b$ all as
$a_1/\cdots/a_k$ with $k=2$.
When ! appears in front of a more complex expression (as in $!(a|b)$),
we treat it separately. We only found 10 expressions that use $!$ and
are different from the expression $!a$.

Furthermore, each row represents the expression type listed on the left plus its
symmetric form. For instance, when we write $a^*/b$, we count the
expressions of the form $a^*/b$ and $b/a^*$. The variant listed in the
table is the one that occurred most often in the data. That is,
$a^*/b$ occurred more often than $b/a^*$.

Bagan et al.~\cite{BaganBG-pods13} proved a dichotomy on the data complexity of
evaluating property paths under a \emph{simple path} semantics, i.e.,
expressions can only be matched on paths in the RDF graph in which
nodes appear only once. They showed that, although evaluating property
paths under this semantics is \np-complete in general, it is possible
in \ptime if the expressions belong to a class called
$C_\text{tract}$. Remarkably, we only found one expression in our
corpus which is not in $C_\text{tract}$, namely $(a/b)^*$.

\begin{table}[tb]
  \centering
  \begin{tabular}{@{}crrr@{}}
    \toprule
    \emph{Expression Type} & \emph{Absolute} & \emph{Relative} & $k$\\
    \midrule
    $(a_1| \cdots |a_k)^*$ & 72,009 & 39.12\% & 2--4\\
    $a^*$ & 48,636 & 26.42\%\\
    $a_1/\cdots/a_k$ & 21,435 & 11.65\% &  2--6\\
    $a^*/b$ & 19,126 & 10.39\%\\
    $a_1| \cdots |a_k$ & 16,053 & 8.72\% & 2--6\\
    $a^+$ & 3,805 & 2.07\%\\
    $a_1?/\cdots/a_k?$ & 2,855 & 1.55\% & 1--5\\
    $a(b_1|\cdots|b_k)$ & 37 & 0.02\% & 2\\
    $a_1/a_2?/\cdots/a_k?$ & 31 & 0.02\% & 1--3\\
    $(a/b^*)|c$ & 15 & 0.01\%\\
    $a^*/b?$ & 13 & 0.01\% \\
    $a/b/c^*$ & 11 &  0.01\%\\
    $!(a|b)$ & 10 & 0.01\%\\
    $(a_1|\cdots|a_k)^+$ & 10 & 0.01\% & 2\\
    $(a_1|\cdots|a_k) (a_1|\cdots|a_k)$ & 5 & $<$ 0.01\% & 2--6\\
    $a?|b$ & 2 & $<$ 0.01\%\\
    $a^*|b$ & 2 & $<$ 0.01\%\\
    $(a|b)?$ & 2 & $<$ 0.01\%\\
    $a|b^+$  & 1 & $<$ 0.01\%\\
    $a^+|b^+$ & 1 & $<$ 0.01\%\\
    $(a/b)^*$ & 1 & $<$ 0.01\% \\
\end{tabular}
  \caption{Structure of navigational property paths in our corpus \label{fig:propertypaths2}}
\end{table}

 \section{Evolution of Queries over Time}\label{sec:streaks}

In a typical usage scenario of a SPARQL endpoint, a user queries the
data and gradually refines her query until the desired result is
obtained. In this section, we analyse to which extent such behavior
occurs. The results are very preliminary but show that, in certain
contexts, it be interesting to investigate optimization techniques for
sequences of similar queries.

We consider a query log to be an ordered list of queries $q_1, \ldots,
q_n$. We introduce the notion of a \emph{streak}, which intuitively
captures a sequence of similar queries within close distance of each
other. To this end we assume the existence of a \emph{similarity test}
between two queries.
We then say that queries $q_i$ and
$q_j$ with $i < j$ \emph{match} if (1) $q_i$ and $q_j$ are similar and
(2) no query $q_{i'}$ with $i < i' < j$ is similar to $q_i$.  A
\emph{streak (with window size $w$)} is a sequence of queries
$q_{i_1}, \ldots, q_{i_k}$ such that, for each $\ell = 1,\ldots,k-1$,
we have that $i_{\ell+1} - i_\ell \leq w$ and $q_{i_{\ell+1}}$ matches
$q_{i_\ell}$.

In theory, it is possible for a query to belong to multiple
streaks. E.g., it is possible that $q_1$ and $q_2$ do not match, but
query $q_3$ is sufficiently similar to both. In this case, $q_3$
belongs to both streaks starting with $q_1$ and with $q_2$.

In the present study, we used Levenshtein distance as a similarity
test. More precisely, we said that two queries are \emph{similar} if
their Levenshtein distance, after removal of namespace prefixes, is at
most 25\%.\footnote{We normalized the measure by dividing the Levenshtein
  distance by the length of the longer string.} We removed namespace prefixes prior to measuring their
Levenshtein distance, because they introduce superficial
similarity. As such, we require queries to be at least 75\% identical
starting from the first occurrence of the keywords \select, \ask,
\construct, or \describe. We took a window size of 30.

Since the discovery of streaks was extremely resource-consuming, we
only analysed streaks in randomly selected log files from
\dbpediaFour, \dbpediaFift, and \dbpediaSixt.  The sizes of these
three log files, each reflecting a single day of queries to the
endpoint, were 273MiB, 803MiB, and 1004MiB respectively.
For the ordering of the queries, we simply considered
the ordering in the log files, since the logs are sorted over time.

\begin{table}[tb]
  \centering
  \begin{tabular}{@{}r rrrr@{}}
    \toprule
    \emph{Streak length}  & \emph{\#DBP'14} & \emph{\#DBP'15} & \emph{\#DBP'16}\\
    \midrule
    1--10 & 42,272 & 167,292 & 199,375\\
    11--20 & 3,732 & 24,001 & 37,402\\
    21--30 & 2,425 & 4,813 & 17,749\\
    31--40 & 884 & 667 & 5,849\\
    41--50 & 283 & 162 & 1,998\\
    51--60 & 88 & 40 & 711\\
    61--70 & 26 & 8 & 357\\
    71--80 & 15 & 4 & 129\\
    81--90 & 5 & 1 & 54\\
    91--100 & 4 & 0 & 27\\
    $>$100 & 5 & 0 & 24 \\
  \end{tabular}
  \caption{Length of streaks in three single-day log files}
\end{table}

Using window size 30, the longest streak we found had length 169 and
was in the 2016 log file. When we increased the window size, we
noticed that it was still possible to obtain longer streaks.
We believe that a more refined analysis on the encountered streaks can be
carried out when tuning the window size and deriving more complex metrics
on the similarity of the queries within each streak.
These issues are, however, subject of further research, which we plan to
pursue in future work.

 \section{Conclusions and Discussion}

We have conducted an extensive analytical study on a large corpus of real SPARQL query logs.
Our corpus is inherently heterogeneous and consists of a majority of DBpedia query logs along with query logs on biological
datasets (namely BioPortal and BioMed datasets) and geological data\-sets (LGD), query logs on bibliographic data (SWDF),
and query logs from a museum SPARQL endpoints (British Museum).  We have
completed this corpus with the example queries from Wikidata (Feb.\ 2017), which
are cherry picked from real SPARQL queries on this data source.
The majority of the datasets
exhibit similar
characteristics, such as for instance the simplicity of queries amounting
to 1 or 2 triples. The only exception occurs with British Museum and
Wikidata datasets (Figure \ref{fig:triplecount}), where the former is a set of queries generated from fixed templates and the latter is a query \emph{wiki} rather than a query log. Clearly, the DBpedia datasets
are the most voluminous and recent in our corpus, thus making their results
quite significant. For instance, despite the fact that single triple queries are
numerous in these datasets, more complex queries (with 11 triples or more)
have lots of occurrences (up to 21\% of the total number of queries
for DBpedia13).
Moreover, we observed that most of the analyzed queries across all datasets are \select/\ask queries,
which range between 91\% and 99.88\% for all datasets except DBpedia16 and LGD13, that have lower percentages.
Therefore, we focused on such queries in the remainder of the paper since
these queries turn out to be the queries that users most often formulate in
SPARQL query endpoints. We have further examined the occurrences of operator distributions and the number of projections and subqueries.
This analysis lets us address a specific fragment, namely the
\and/\opt/\filter patterns (\aof patterns).
For such patterns, we derived the graph- and hypergraph structures and
analyzed the impact of the structure on query evaluation. We simulated real
chain and cycle query logs with a synthetic generator by building diverse workloads of \ask queries and measured their average runtime in two systems, Blazegraph, used by the Wikimedia foundation, and PostgreSQL. In both systems, the difference between average performances of such different query shapes are perceivable. We decided to dig deeper in the shape analysis in order to classify these queries under general query shapes as canonical graphs and characterize their tree-likeness as hypergraphs.
We believe that this shape analysis can serve the need of fostering the discussion on the design of new query languages for
graph data, as pursued by the LDBC Graph Query Language Task Force
\cite{ldbc,FletcherVY17,BarceloDB17}. It can also inspire the
conception of novel query optimization techniques suited for these
query shapes, along with tuning and benchmarking methods. For
instance, we are not aware of existing benchmarks targeting flowers
and flower sets.
The analysis on property paths showed that these are not yet widely used in
the entire corpus, even though they are numerous in the Wikidata corpus. A
recent discussion (July 6th, 2017) in a Neo4J working group
\cite{CypherOcig} concerned the support of full-fledged regular path
queries in OpenCypher. This discussion, and other discussions on
standard graph query languages \cite{ldbc,FletcherVY17,BarceloDB17} could benefit from our analysis, devoted to
find which property paths are actually used most often when ordinary
users have the power of regular expressions.
Finally, we performed an study on the way users specify their
queries in SPARQL query logs, by identifying streaks of similar queries.
This analysis is for instance crucial to understand query specification from real
users and thus usability of databases, which is an hot research topic in our community \cite{JagadishCEJLNY07, NandiJ11}.
Our analysis has been carried out with scripts in different languages,
amounting to a total of roughly $9,000$ source lines of code (SLOC).
We plan to make
these scripts publicly available in the next months.

\balance

\section*{Acknowledgments}
We would like to acknowledge USEWOD and Patrick van Kleef and the entire
team of OpenLink
Software
for hosting the
official DBPedia endpoint and granting us the access to the large DBPedia
query logs analysed in
this paper.

\bibliographystyle{abbrv}

\newpage

\begin{appendix}

The appendix contains the results of our analytical study when applied to
the larger set of \emph{Valid} queries containing duplicates. The
characteristics of this corpus containing a total of 173,798,237 queries are shown in Table~\ref{tab:datasets} in the body of the paper.

\begin{table}[tb]
  \centering
\begin{tabular}{@{}r rr@{}}
  \toprule
  \emph{Element}  & \emph{Absolute} & \emph{Relative}\\
  \midrule
  \select   & 160,722,786 & 31.10\%\\
  \ask      &   4,680,967 &  0.91\%\\
  \describe &   7,127,250 &  1.38\%\\
  \construct&   1,916,852 &  0.37\%\\
  \midrule
  \distinct &  53,440,345 & 10.34\%\\
  \limit    &  24,964,363 &  4.83\%\\
  \offset   &   6,646,757 &  1.29\%\\
  \orderby  &   2,727,496 &  0.53\%\\
  \midrule
  \filter   &  73,055,654 & 14.13\%\\
  \and      &  61,417,138 & 11.88\%\\
  \union    &  36,585,529 &  7.08\%\\
  \opt      &  52,145,320 & 10.09\%\\
  \graph    &  27,514,010 &  5.32\%\\
  \notexists&   1,889,531 &  0.37\%\\
  \minus    &   1,281,221 &  0.25\%\\
  \exists   &      11,139 & $<$ 0.00\%\\
  \midrule
  \countop  &     373,906 &  0.07\%\\
  \maxop    &       6,212 &  $<$0.00\%\\
  \minop    &       6,833 &  $<$0.00\%\\
  \avgop    &       2,993 &  $<$0.00\%\\
  \sumop    &         392 &  $<$0.00\%\\
  \groupby  &     329,226 &  $<$0.06\% \\
  \having   &      20,415 &  $<$0.00\%\\
\end{tabular}
\caption{Keyword count in queries\label{tab:keywords:complete}}
\end{table}

Precisely, in the order of appearance, we have repeated the shallow
analysis on this corpus and obtained the keyword count in queries in
Table~\ref{tab:keywords:complete}, along with the operator
distribution in Table~\ref{tab:operators2:complete}.
The percentages of queries exhibiting different number of triples for this
complete corpus are reported in Figure~\ref{fig:triplecount:complete}.

\begin{table}[tb]
  \centering
  \begin{tabular}{@{}r <{\kern\tabcolsep} @{} R r@{}}
    \toprule
    \emph{Operator Set} & \emph{Absolute} & \emph{Relative}\\
    \midrule
    \emph{none}                                    &  42,012,743 &  25.40\%\\
    \textsf{F}                                     &  16,155,263 &   9.77\%\\
    \textsf{A}                                     &   8,055,974 &   4.87\%\\
    \textsf{A}, \textsf{F}                         &   6,830,892 &   4.13\%\\
    \rowcolor{black!10}[0pt][0pt] CPF subtotal     &  73,054,872 &  44.17\%\\
    \midrule
    \textsf{O}                                     &   2,743,584 &   1.66\%\\
    \textsf{O}, \textsf{F}                         &   3,400,506 &   2.06\%\\
    \textsf{A}, \textsf{O}                         &   6,096,091 &   3.69\%\\
    \textsf{A}, \textsf{O}, \textsf{F}             &  13,612,119 &   8.23\%\\
    \rowcolor{black!10}[0pt][0pt] CPF+\textsf{O}   & +25,852,257 & +15.63\%\\
    \midrule
    \textsf{G}                                     &  26,288,134 &  15.89\%\\
    \rowcolor{black!10}[0pt][0pt] CPF+\textsf{G}   & +26,288,951 & +16.15\%\\
    \midrule
    \textsf{U}                                     &   7,267,329 &   4.39\%\\
    \textsf{U}, \textsf{F}                         &     567,912 &   0.34\%\\
    \textsf{A}, \textsf{U}                         &   1,102,282 &   0.67\%\\
    \textsf{A}, \textsf{U}, \textsf{F}             &   1,416,960 &   0.86\%\\
    \rowcolor{black!10}[0pt][0pt] CPF+\textsf{U}   & +10,354,483 &  +6.26\%\\
    \midrule
    \textsf{A}, \textsf{O}, \textsf{U}, \textsf{F} &  24,520,317 &  14.82\%\\
  \end{tabular}
  \caption{Sets of operators used in queries: \filter (\textsf{F}), \and (\textsf{A}), \opt
    (\textsf{O}), \graph (\textsf{G}), and \union (\textsf{U})}
    \label{tab:operators2:complete}
\end{table}

\begin{figure*}[p!]
  \centering

\pgfplotstableread[col sep=comma]
{appendix2_triples_sa_all.csv}
{\datatableF}
\begin{tikzpicture}
\begin{axis}[
  ybar stacked,
    legend style={at={(1,0.5)}, anchor=west},
    reverse legend,
    ylabel near ticks,
    yticklabel=\pgfmathparse{100*\tick}\pgfmathprintnumber{\pgfmathresult}\,\%,
    bar width=0.7cm,
    x=1.0cm,
    xticklabels={
      \dbpediaTwel,
      \dbpediaThir,
      \dbpediaFour,
      \dbpediaFift,
      \dbpediaSixt,
      \lgdThir,
      \lgdFour,
      \biopThir,
      \biopFour,
      \biomedThir,
      \swdfThir,
      \rkbeFour,
      \wikiSeven
    },
    xtick={1, ..., 13},
    x tick label style={rotate=45,anchor=east},
    legend cell align=left
  ]
  \addplot+[NavyBlue!80!black,fill=NavyBlue,ybar] table[x=id, y=0] from {\datatableF};
  \addplot+[ProcessBlue!80!black,fill=ProcessBlue,ybar] table[x=id, y=1] from {\datatableF};
  \addplot+[SkyBlue!80!black,fill=SkyBlue,ybar] table[x=id, y=2] from {\datatableF};
  \addplot+[BlueGreen!80!black,fill=BlueGreen,ybar] table[x=id, y=3] from {\datatableF};
  \addplot+[LimeGreen!80!black,fill=LimeGreen,ybar] table[x=id, y=4] from {\datatableF};
  \addplot+[SpringGreen!80!black,fill=SpringGreen] table[x=id, y=5] from {\datatableF};
  \addplot+[Goldenrod!80!black,fill=Goldenrod] table[x=id, y=6] from {\datatableF};
  \addplot+[YellowOrange!80!black,fill=YellowOrange,ybar] table[x=id, y=7] from {\datatableF};
  \addplot+[Orange!80!black,fill=Orange,ybar] table[x=id, y=8] from {\datatableF};
  \addplot+[Red!50!black,fill=Red,ybar] table[x=id, y=9] from {\datatableF};
  \addplot+[BrickRed!50!black,fill=BrickRed,ybar] table[x=id, y=10] from {\datatableF};
  \addplot+[black,fill=Sepia,ybar] table[x=id, y=11] from {\datatableF};

  \legend{0, 1, 2, 3, 4, 5, 6, 7, 8, 9, 10, $11+$}
\end{axis}
\end{tikzpicture}

\centering
\resizebox{\textwidth}{!}{
\begin{tabular}{@{}cccccccccccccc@{}}
\toprule
Datasets &  \dbpediaTwel & \dbpediaThir  &   \dbpediaFour &   \dbpediaFift &   \dbpediaSixt &  \lgdThir   &     \lgdFour  &     \biopThir &    \biopFour &     \biomedThir  &     \swdfThir   &     \rkbeFour   &    \wikiSeven  \\
\textit{S/A} & 99.15\% & 91.88\% & 95.38\% & 93.05\% & 63.99\% &
29.01\% & 97.47\% & 100\% & 99.69\% & 12.87\% & 96.14\% & 98.64\% & 99.68\%\\
\textit{Avg}$_{\#T}$ & 2.38 & 3.98  &  2.09 &  2.94 &   3.78 &  3.19   &   2.65  &  1.16 &  1.42 &   2.44  &    1.51   &     5.47   &    3.94  \\
\bottomrule
\end{tabular}
}
\caption{Percentages of queries exhibiting different number of triples
  (in colors) for each dataset (top), the average numbers of triples
  of the queries for each dataset (\textit{Avg}$_{\#T}$, bottom), and
  the percentage of \select/\ask-queries (\textit{S/A},
  bottom).}\label{fig:triplecount:complete}
\end{figure*}

Figure~\ref{fig:cq-size:complete} shows the relative sizes of the different
fragments of conjunctive queries, namely
$CQ$, $CQ_F$ and $CQ_{OF}$.

\begin{figure}[p!]
  \centering
  \pgfplotstableread[col sep=comma]
  {appendix2_triples-cqs_all.csv}
  {\datatableF}

  \begin{tikzpicture}
    \begin{axis}[
      enlarge x limits=0.5,
      ybar stacked,
      legend style={at={(1,0.5)}, anchor=west},
      reverse legend,
      ylabel near ticks,
      yticklabel=\pgfmathparse{100*\tick}\pgfmathprintnumber{\pgfmathresult}\,\%,
      bar width=0.7cm,
      x=1cm,
      xticklabels={
        \cq,
        \cqf,
        \cqof
      },
      xtick={1, ..., 3},
      x tick label style={rotate=45,anchor=east},
      legend cell align=left
      ]
       \addplot+[NavyBlue!80!black,fill=NavyBlue,ybar] table[x=id, y=0] from {\datatableF};
      \addplot+[ProcessBlue!80!black,fill=ProcessBlue,ybar] table[x=id, y=1] from {\datatableF};
      \addplot+[SkyBlue!80!black,fill=SkyBlue,ybar] table[x=id, y=2] from {\datatableF};
      \addplot+[BlueGreen!80!black,fill=BlueGreen,ybar] table[x=id, y=3] from {\datatableF};
      \addplot+[LimeGreen!80!black,fill=LimeGreen,ybar] table[x=id, y=4] from {\datatableF};
      \addplot+[SpringGreen!80!black,fill=SpringGreen] table[x=id, y=5] from {\datatableF};
      \addplot+[Goldenrod!80!black,fill=Goldenrod] table[x=id, y=6] from {\datatableF};
      \addplot+[YellowOrange!80!black,fill=YellowOrange,ybar] table[x=id, y=7] from {\datatableF};
      \addplot+[Orange!80!black,fill=Orange,ybar] table[x=id, y=8] from {\datatableF};
      \addplot+[Red!50!black,fill=Red,ybar] table[x=id, y=9] from {\datatableF};
      \addplot+[BrickRed!50!black,fill=BrickRed,ybar] table[x=id, y=10] from {\datatableF};
      \addplot+[black,fill=Sepia,ybar] table[x=id, y=11] from {\datatableF};

      \legend{2, 3, 4, 5, 6, 7, 8, 9, 10, $11+$}
    \end{axis}
\end{tikzpicture}

  \caption{Size of \cq-like queries with at least two
    triples. \label{fig:cq-size:complete}}
\end{figure}

\begin{table*}[p!]
  \centering
  \begin{tabular}{@{}r <{\kern\tabcolsep} @{} R r@{}}
    & \cq \\
    \toprule
    \emph{Shape} &   \emph{$\#$Queries} &  \emph{Relative $\%$}\\
    \midrule
    single edge                         & 32,980,584 &  82.79\%\\
    chain                               & 39,200,135 &  98.40\%\\
    chain set                           & 39,281,219 &  98.60\%\\
    star                                &    494,071 &   1.24\%\\
    tree                                & 39,711,504 &  99.68\%\\
    forest                              & 39,793,015 &  99.89\%\\
    \midrule
    cycle                               &     39,412 &   0.10\%\\
    flower                              & 39,755,202 &  99.79\%\\
    flower set                          & 39,836,742 &  99.99\%\\
    treewidth $\leq 2$                  & 39,838,786 & 100.00\%\\
    \midrule
    treewidth $= 3$                     &          2 &   0.00\%\\
    \rowcolor{black!10}[0pt][0pt] total & 39,838,788 & 100.00\%\\
  \end{tabular}
  \quad
  \begin{tabular}{@{}r <{\kern\tabcolsep} @{} r@{}}
    \multicolumn{2}{c}{\cqf}\\
    \toprule
    \emph{$\#$Queries} &  \emph{Relative $\%$}\\
    \midrule
                                          46,638,936 &  81.31\%\\
                                          55,286,105 &  96.38\%\\
                                          55,374,432 &  96.54\%\\
                                           1,902,267 &   3.32\%\\
                                          57,216,983 &  99.75\%\\
                                          57,306,126 &  99.91\%\\
    \midrule
                                              39,635 &   0.07\%\\
                                          57,262,849 &  99.83\%\\
                                          57,352,028 &  99.99\%\\
                                          57,360,489 & 100.00\%\\
    \midrule
                                                   2 &   0.00\%\\
    \rowcolor{black!10}[0pt][0pt]         57,360,491 & 100.00\%\\
  \end{tabular}
  \quad
  \begin{tabular}{@{}r <{\kern\tabcolsep}@{} r@{}}
    \multicolumn{2}{c}{\cqof}\\
    \toprule
    \emph{$\#$Queries} &  \emph{Relative $\%$}\\
    \midrule
                                          48,299,192 &  70.41\%\\
                                          61,926,151 &  90.27\%\\
                                          62,057,865 &  90.46\%\\
                                           5,729,035 &   8.35\%\\
                                          68,005,133 &  99.13\%\\
                                          68,140,016 &  99.33\%\\
    \midrule
                                              39,675 &   0.06\%\\
                                          68,058,458 &  99.21\%\\
                                          68,193,382 &  99.41\%\\
                                          68,600,301 & 100.00\%\\
    \midrule
                                                   2 &   0.00\%\\
    \rowcolor{black!10}[0pt][0pt]         68,600,303 & 100.00\%\\
  \end{tabular}
\vspace{-0.18cm}
\caption{Cumulative shape analysis of $CQ$, $CQ_{F}$, $CQ_{FO}$ across
  all logs.
}
\label{tab:shape:complete} \label{tab:shape-noConsts:complete}
\end{table*}

The results of the shape analysis applied to this complete corpus are reported in Table~\ref{tab:shape-noConsts:complete}.

Compared to the \emph{Unique} dataset, reported in the body of the paper,
we can notice that the larger and more complex queries seem to occur more
often in the set with duplicates than in the set without duplicates.

\begin{figure}[htbp]
  \centering
  \begin{tabular}{@{}crrr@{}}
    \toprule
    \emph{Expression Type} & \emph{Absolute} & \emph{Relative} & $k$\\
    \midrule
    $(a_1| \cdots |a_k)^*$              & 274,963 &    55.45\% & 2--4\\
    $a^*$                               &  87,486 &    17.64\%\\
    $a_1/\cdots/a_k$                    &  76,412 &    15.41\% & 2--6\\
    $a^*/b$                             &  19,593 &     3.95\%\\
    $a_1| \cdots |a_k$                  &  18,194 &     3.67\% & 2--6\\
    $a^+$                               &  10,473 &     2.11\%\\
    $a_1?/\cdots/a_k?$                  &   8,511 &     1.72\% & 1--5\\
    $(a/b^*)|c$                         &      45 &     0.01\%\\
    $a(b_1|\cdots|b_k)$                 &      43 &     0.01\% & 2\\
    $a_1/a_2?/\cdots/a_k?$              &      37 &     0.01\% & 1--3\\
    $a^*/b?$                            &      30 &     0.01\% \\
    $a/b/c^*$                           &      14 & $<$ 0.01\%\\
    $!(a|b)$                            &      10 & $<$ 0.01\%\\
    $(a_1|\cdots|a_k)^+$                &      11 & $<$ 0.01\% & 2\\
    $(a_1|\cdots|a_k) (a_1|\cdots|a_k)$ &       8 & $<$ 0.01\% & 2--6\\
    $a?|b$                              &       2 & $<$ 0.01\%\\
    $a^*|b$                             &       2 & $<$ 0.01\%\\
    $(a|b)?$                            &       1 & $<$ 0.01\%\\
    $a|b^+$                             &       1 & $<$ 0.01\%\\
    $a^+|b^+$                           &       1 & $<$ 0.01\%\\
    $(a/b)^*$                           &       1 & $<$ 0.01\% \\
\end{tabular}
\caption{Structure of navigational property paths in our complete corpus\label{fig:propertypaths2:complete}}
\end{figure}

Finally, property paths for the complete corpus containing duplicates
are reported in Figure~\ref{fig:propertypaths2:complete}.

\end{appendix}

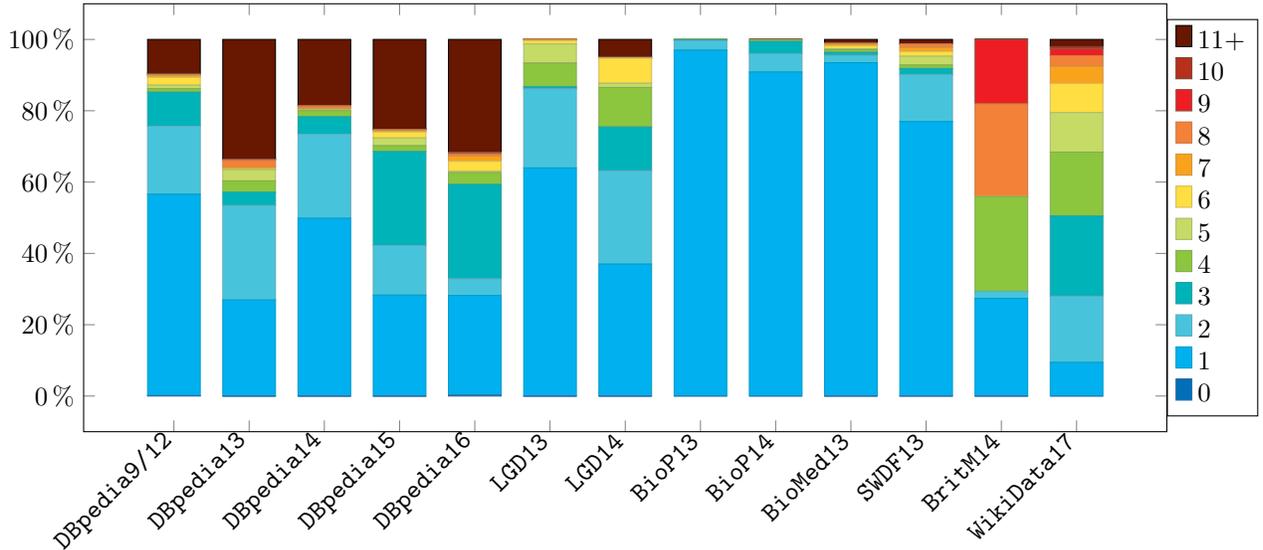
\end{document}